\documentclass[acmsmall,screen,authorversion]{acmart}

\newif\ifdraft
  \drafttrue

  \newif\iflong
    \longtrue

    \graphicspath{ {./images/} }

    \usepackage{mathpartir}
    \usepackage{amsmath}
    \usepackage{mathtools}
    \usepackage[altpo]{backnaur}
    \usepackage{algorithm2e}
    \usepackage{xspace}
    \usepackage{stmaryrd}
    \usepackage[export]{adjustbox}
    \usepackage[capitalize]{cleveref}
    \usepackage{enumitem}
    \usepackage{etoolbox}
    \usepackage{thm-restate}

    \usepackage{booktabs}

    \DeclareMathOperator{\ret}{\mathrm{ret}}
    \DeclareMathOperator{\d=}{\ \triangleq\ }

    \DeclareMathOperator{\agt}{\mathcal{A}}
    \DeclareMathOperator{\ops}{\mathcal{O}}
    \DeclareMathOperator{\val}{\mathcal{V}}
    \DeclareMathOperator{\var}{\mathcal{X}}
    \DeclareMathOperator{\typ}{\mathcal{T}}
    \DeclareMathOperator{\emap}{G}
    \DeclareMathOperator{\interp}{\mu}
    \DeclareMathOperator{\M}{M}

    \newcommand{\vset}{\overline{V}}

    \DeclareMathOperator{\f}{\mathtt{Form}}
    
    \DeclareMathOperator{\term}{\mathtt{Term}}
    \DeclareMathOperator{\ef}{\alpha}
    
    \DeclareMathOperator{\dom}{\text{dom}}
    \DeclareMathOperator{\fv}{\mathtt{FV}}
    \newcommand{\emb}[1]{\widehat{#1}}
    \newcommand{\vemb}[1]{\widehat{#1}}
    
    \renewcommand{\exp}{\mathtt{Exp}}
    \newcommand{\n}[1]{N(#1)}
    \renewcommand{\v}[0]{\text{Val}}
    \newcommand{\ite}[3]
    {%
      \ifstrempty{#1}{%
        \text{\text{ite}}
      }{%
        {\text{ite}(#1, #2, #3)}
      }%
    }

    \newcommand{\BB}{\mathbb{B}}
    \newcommand{\II}{\mathbb{I}}
    \newcommand{\NN}{\mathbb{N}}
    
    \newcommand{\RR}{\mathbb{R}}

    \newcommand{\sem}[1]{\llbracket {#1} \rrbracket}
    \newcommand{\rname}[1]{[\textsc{#1}]}

    \newcommand{\SYSTEM}{\textsc{Slice}\xspace}
    \newcommand{\kw}[1]{\mathsf{#1}}
    \newcommand{\kwtrue}{\kw{true}}
    \newcommand{\kwfalse}{\kw{false}}
    \newcommand{\kwlet}{\kw{let}}
    \newcommand{\kwin}{\kw{in}}
    \newcommand{\kwif}{\kw{if}}
    \newcommand{\kwthen}{\kw{then}}
    \newcommand{\kwelse}{\kw{else}}
    \newcommand{\kwlt}{\kw{left}}
    \newcommand{\kwrt}{\kw{right}}
    \newcommand{\kwcomp}{\kw{compare}}
    \newcommand{\kwsplit}{\kw{split}}
    \newcommand{\kwdiv}{\kw{divide}}
    \newcommand{\kwmark}{\kw{mark}}
    \newcommand{\kweval}{\kw{eval}}
    
    \newcommand{\kwcake}{\kw{cake}}
    \newcommand{\kwinit}{\kwcake}
    \newcommand{\kwdef}{\kw{def}}
    \newcommand{\kwalloc}{\kw{alloc}}
    \newcommand{\kwsort}{\kw{sort}}

    \newcommand{\ediv}[2]{\kwdiv({#1}, {#2})}
    \newcommand{\emark}[3]{\kwmark_{{#1}}({#2}, {#3})}
    \newcommand{\eeval}[2]{\kweval_{{#1}}({#2})}

\newcommand{\append}[1]{\iflong%
\Cref{#1}\else%
the supplementary materials\fi%
}

\AtBeginDocument{%
  }

\setcopyright{rightsretained}
\acmJournal{PACMPL}
\acmYear{2023}
\acmVolume{7}
\acmNumber{PLDI}
\acmArticle{179}
\acmMonth{6}
\acmPrice{}
\acmDOI{10.1145/3591293} 
\copyrightyear{2023}
\acmSubmissionID{pldi23main-p612-p}
\received{2022-11-10}
\received[accepted]{2023-03-31}

\begin{document}
\title{Cutting the Cake: A Language for Fair Division}


\author{Noah Bertram}
\affiliation{%
  \institution{Cornell University}
  \country{USA}
}

\author{Alex Levinson}
\affiliation{%
  \institution{Cornell University}
  \country{USA}
}

\author{Justin Hsu}
\affiliation{%
  \institution{Cornell University}
  \country{USA}
}



\begin{abstract}
  The \emph{fair division} literature in economics considers how to divide
  resources between multiple agents such that the allocation is
  \emph{envy-free}: each agent receives their favorite piece. Researchers have
  developed a variety of fair division protocols for the most standard setting,
  where the agents want to split a single item, however, the protocols are highly
  intricate and the proofs of envy-freeness involve tedious case analysis.

  We propose \SYSTEM, a domain specific language for fair-division. Programs in
  our language can be converted to logical formulas encoding envy-freeness and
  other target properties. Then, the constraints can be dispatched to automated
  solvers. We prove that our constraint generation procedure is sound and
  complete. We also report on a prototype implementation of \SYSTEM, which we
  have used to automatically check envy-freeness for several protocols from the
  fair division literature.
\end{abstract}

\ccsdesc[500]{Note sure}
\ccsdesc[300]{Also not sure}
\ccsdesc{also also not sure}
\ccsdesc[100]{also also also not sure}

\keywords{Fair division, automated verification}

\maketitle
\section{Introduction}
\label{sec:intro}

Suppose we want to divide a valuable item between a set of agents. How can we
ensure that the division is fair? Ideally, we would divide the item into equal
pieces and give one piece to each agent. However, in reality, agents often have
different preferences and may disagree about whether an item has been divided
into equal pieces or not. What does it mean for an item to be divided fairly?

In economics, the field of \emph{fair division} studies these kinds of
questions. Researchers have proposed different definitions of fairness, and
considered whether a fair allocation is possible in a variety of settings---the
item might be divisible or indivisible; there may be geometric constraints
depending on the shape of the item, or limits on which kinds of splits are
allowed.

We focus on the most well-studied model, called \emph{fair cake-cutting}. Here,
there is a single item---a ``cake''---that is infinitely divisible: it can be
cut into arbitrarily small pieces. Each agent has a \emph{valuation} function
that assigns numeric values to subsets of the cake. The goal is to divide the
cake into disjoint pieces and allocate one piece to each agent, such that each
agent does not prefer any other agent's piece to their own. Such an allocation
is called \emph{envy-free}.

\subsection{Protocols for Fair Cake-Cutting: Challenges and Complexity}

While the existence of a fair allocation is often easy to show, assuming mild
conditions on the agent valuations, computing a fair allocation is more
challenging. Algorithms for fair division first specify a sequence of cuts that
agents should make, depending on how they value the current pieces. Then,
protocols describe how to allocate the resulting pieces to agents; for instance,
agents might select their favorite piece in some fixed order, or one agent may
be required to take a particular piece.

For a simple example, suppose we want to divide a slice of cake between two
children so that each child does not envy the other's piece. First, we can ask
one child to split the slice into two pieces that they consider to be equal.
Then, we can ask the other child to pick their preferred piece. This simple
protocol, also known as \emph{Cut-Choose} or Divide-and-Choose, produces an
envy-free allocation: as long as both children follow the protocol, they will
not wish they received the other's piece.

While this two-agent protocol is easy to describe and justify, the situation
becomes much more difficult with more agents. An envy-free protocol for three
agents was not known until 1960~\citep{robertson1998cake}, and finding a bounded
envy-free protocol for four agents was an open problem until 2016, solved
by~\citet{aziz2016discrete}; their procedure can make up to 203 cuts. A protocol
for any number of agents soon followed~\citep{aziz2016discreteany}, but it is
tremendously complex---the number of cuts is finite, but bounded by a tower of
five exponentials in $n$, where $n$ is the number of agents.

Although such envy-free protocols are impressive achievements, they are not easy
to implement. The protocols are typically specified in lengthy pseudocode, often
interspersed with informal English. Proving the key envy-freeness property is a
highly tedious task that involves considering a large number of cases. For most
protocols, it is not feasible to spell out all of the details in the proof.
Besides violating envy-freeness, there are other ways a protocol could go wrong.
For instance, a protocol could ask an agent to cut past the end of piece, or
allocate the same part of a cake more than once.

\subsection{Our Work: Automatically Verifying Cake-Cutting Protocols}

We consider how to \emph{formally verify} envy-freeness for fair division
protocols. Our hypothesis is that since proofs of this property typically
require tedious, but fairly straightforward case analyses, they may
be a good target for automated solvers.

Concretely, we take a language-based approach. First, we develop a core language
for cake cutting protocols. Our language can capture all protocols in the
Robertson-Webb query model~\citep{robertson1998cake}, a standard computational
model in the fair division literature. Our operational semantics allows
\emph{uncountable non-determinism}---required for situations when an agent may
consider multiple pieces to be equally good---and also models protocol errors,
like cutting out of bounds. Expressing the protocols in our language removes any
ambiguity from their pseudocode descriptions.

Second, we develop a constraint generation procedure for our language. At a high
level, we translate the operational behavior of our programs into first-order
logic formulas, parametrized by the unknown valuation functions. We prove
that our constraint translation is sound and complete, and demonstrate how to
use the constraints to encode properties like envy-freeness.


Finally, we implement our language in a prototype tool called \SYSTEM. Users can
write cake-cutting protocols in our language, which provides primitive
operations for fair division. Our tool converts programs to constraints
guaranteeing envy-freeness, which are then dispatched to an SMT solver. Our
implementation also includes an evaluator, which can directly execute the
verified protocols. We evaluate our tool by implementing several cake-cutting
protocols and automatically verifying envy-freeness. As far as we are aware, our
method is the first to formally verify these protocols and properties.


\paragraph{Outline}
After introducing preliminaries about fair cake-cutting and illustrating our
approach on a simple example (\Cref{sec:motiv}), we present our primary
technical contributions:
\begin{itemize}
  \item We develop a core language for fair cake-cutting protocols, based on the
        Robertson-Webb query model (\Cref{sec:language}).
  \item We design a constraint translation from our core language to formulas in
        first-order logic.
        We prove that our translation is sound and complete, and we show how
        to use our constraints to encode envy-freeness and other target properties
        (\Cref{sec:constraints}).
  \item We implement several proposed protocols from the cake-cutting literature
        in our language (\Cref{sec:examples}).
  \item We develop a prototype implementation for writing, verifying, and
        executing protocols. Our tool can verify
        envy-freeness for example protocols automatically (\Cref{sec:impl}).
\end{itemize}
We survey related work in \Cref{sec:rw} and discuss future directions in
\Cref{sec:conc}.

\section{Verifying Fair Division: Preliminaries and a Toy Example}
\label{sec:motiv}

Fair division is a rich and well-studied area of
economics~\citep{robertson1998cake}. We focus on the most common setting:
\emph{fair cake-cutting}.

\subsection{Mathematical Preliminaries}

\paragraph*{Cakes and pieces.} We suppose that the agents are interested in
dividing an infinitely divisible item, called the \emph{cake}, which is modeled
as the closed unit interval $[0, 1]$. Each agent receives a subset of the cake,
where the set $P$ of allowed subsets consists of finite unions of disjoint
intervals:
\[
  P \triangleq \{
  \langle a_1, b_1 \rangle \uplus \cdots \uplus \langle a_{k},b_{k} \rangle
  \mid 0 \leq a_i \leq b_i \leq 1, k \in \mathbb{N} \},
\]
Each interval can be open or closed at either endpoint, i.e., $\langle a, b
\rangle$ stands for one of the four sets: $[a,b], (a,b), [a,b), (a,b]$. We refer
to an element of $P$ as a \emph{piece}; note that it it may consist of multiple
disjoint intervals.


\paragraph*{Agents and valuations.} Fair division protocols are designed for a
fixed, finite number of agents, the set of which we will denote with $\mathcal{A}$.
Each agent $a \in \mathcal{A}$ has a \emph{valuation function} $V_a : P \to
  \RR$; agents prefer pieces with larger valuations. The fair division literature
considers various standard assumptions on valuation functions. For concreteness,
we will assume the following four conditions:
\begin{description}
  \item[Normalization.] The value of the whole cake is $1$: $V([0, 1]) = 1$.
  \item[Non-negativity.] The value for any piece $p$ is non-negative: $V(p) \geq 0$.
  \item[Additivity.] If $p_1,p_2 \in P$ are disjoint, then $V(p_1 \cup p_2) = V(p_1) +
      V(p_2)$.
  \item[Continuity.] For $\ell, v \in [0, 1]$, if $v \leq V(\left[ \ell,1 \right])$
    then there exists $r \in [\ell, 1]$ such that $V(\left[ \ell,r \right]) = v$.
\end{description}
These conditions are commonly assumed in the fair division literature. The
first property normalizes the value of the entire cake to be $1$. The second
property means that there is no ``bad'' piece of cake. The third states that
the value of two disjoint pieces is the sum of the values of the pieces.
Combined with the second property, this implies that valuations are
\emph{(weakly) monotonic}: if a piece $p_1$ is contained in another piece $p_2$,
then $V(p_1) \leq V(p_2)$.

The final property states that the value of any piece increases smoothly when
we enlarge a piece, so that there are no sudden jumps in value. Note that the
real number $r$ may not be unique---there could be multiple intervals with the
same value starting at left endpoint $\ell$. Though this property may appear
complicated, it is crucial to support basic operations in cake-cutting
protocols. For instance, it ensures that given an interval $[a, b]$ with value
$v$, we can find a sub-interval $[a, c]$ that has value exactly $v / 2$. By
additivity, the rest of the interval, $(c, b]$, must also have value $v /2$.
In other words, the continuity property ensures that we can always split an
interval into two pieces with equal value.

Another useful consequence of continuity is \emph{atomicity}: $V(\left[ r,r
    \right]) = 0$ for all $r\in [0,1]$. This means we can ignore whether the
endpoints of intervals are closed or open, and we can treat intervals that
overlap only at an endpoint as if they were actually disjoint. For instance,
given $a \leq b \leq c$, we have
\begin{align}
  v(\left[ a,c \right])
   & = v(\left[ a,b \right)) + v(\left[ b, c \right])
  \tag{additivity}                                                \\
   & = v(\left[ a,b \right)) + v([b, b]) + v(\left[ b, c \right])
  \tag{atomicity}                                                 \\
   & = v([a, b]) + v([b, c])
  \tag{additivity}
\end{align}
so additivity continues to hold when two intervals share a single endpoint.

\paragraph*{Allocations and envy-freeness.}
An \emph{allocation} is an
assignment $\{p_{a}\}_{a \in \mathcal{A}}$ of disjoint pieces to agents; agent $a$
receives piece $p_a$. The union of the pieces in an allocation is not required
to be the entire cake---we can leave some parts of the cake unallocated.

The goal of fair division is to arrive at an \emph{envy-free} allocation.
\begin{definition}\label{def:envy}
  An allocation $\{p_{a}\}_{a \in \mathcal{A}}$, is \emph{envy-free} if for all $a,a'\in \mathcal{A}$, $V_{a}(p_{a}) \geq V_{a}(p_{a'})$.
\end{definition}
Intuitively, no agent prefers someone else's piece to their own.  The simplest
example of an envy-free allocation is $p_{a} = \emptyset$ for all $a\in
\mathcal{A}$. This is envy-free as $V_{a}(\emptyset) = 0$ for all $a \in
\mathcal{A}$, since
\[
  V_{a} (\emptyset) = V_{a}(\emptyset \cup \emptyset) = V_{a}(\emptyset) + V_{a}(\emptyset).
\]
While this allocation is envy-free, it is quite wasteful. Fair division
protocols typically aim to allocate most, if not all, of the cake.

\subsection{Cake-Cutting Protocols}

Conceptually, a cake-cutting protocol takes agent valuations as input, and
produces an allocation as output. However, this simple picture is
impractical---valuations assign a numeric value to every possible piece, and it
is unrealistic to assume that agents can communicate their entire valuation
function to the procedure.

Accordingly, algorithms for cake cutting typically only assume indirect access
to the valuation functions through primitive operations, called \emph{queries}.
Operationally, algorithms can be executed as interactive protocols where a
central coordinator queries specific agents to perform actions depending on
their valuation function, and agents are assumed to follow these directions
honestly. Common actions include:
\begin{description}
  \item[Split.] An agent cuts a given piece into two parts that they value equally.
  \item[Trim.] An agent cuts a piece into two parts, typically called the \emph{main
      piece} and a \emph{trimmed piece}, such that the agent is indifferent between
    the main piece and a given reference piece.
  \item[Compare.] An agent reports which of two pieces they prefer more.
  \item[Select.] An agent takes their favorite piece from a given collection of pieces.
\end{description}
Then, a protocol description asks agents to perform actions in some specified
order. Protocols can branch on the result of queries, for instance, following
one sub-protocol if an agent reports that one piece is larger than another, and
a different sub-protocol otherwise. Finally, the pieces that an agent selects
form the final allocation to the agent.

While the operational model appears simple, there are some subtleties. First,
operations can have pre-conditions---if these requirements don't hold, then the
operation is not well-defined. For instance, an action might ask an agent to
trim a piece to have value $v$, but this isn't possible if the entire piece has
value strictly less than $v$. Second, protocol execution can be
non-deterministic. For example, if an agent is asked to divide a piece into two
pieces that they consider to be equally good, there may be multiple possible
divisions because the agent may have zero value for some parts of the cake.
Though the dividing agent may not care about this choice, the choice may affect
the behavior of the other agents, and affect the rest of the protocol. An
envy-free protocol must be envy-free for \emph{all} possible executions.


\subsection{Toy Example: Cut-Choose}
A classical protocol for dividing an item between two agents is
\emph{Cut-Choose}:
\begin{enumerate}
  \item Ask first agent to divide cake into two equal pieces.
  \item Ask second agent to pick their preferred piece.
  \item Ask first agent to pick the remaining piece.
\end{enumerate}
This protocol can be implemented as the following program in a simplified
version of our language; \cref{sec:language} describes our language in full
detail.
\[
  \begin{array}{l}
    \kwlet\ (\mathit{piece}, \mathit{piece}') = \kwsplit_1(\kwcake)\ \kwin     \\
    \kwif\ \kwcomp_2(\mathit{piece}, \mathit{piece}') = \kw{bigger}\ \kwthen \\
    \quad \kwalloc(1 \mapsto \mathit{piece}', 2 \mapsto \mathit{piece})        \\
    \text{else }                                                               \\
    \quad \kwalloc(1 \mapsto \mathit{piece}, 2 \mapsto \mathit{piece}')
  \end{array}
\]
Above, the operations $\kwsplit$ and $\kwcomp$ correspond to the actions
\textbf{Split} and \textbf{Compare} we introduced above. We use the subscripts
$1$ and $2$ on these operations to indicate which agent is asked to perform the
action, i.e., which agent's valuation is queried. We suppose that the comparison
operation $\kwcomp$ returns $\kw{bigger}$ if the agent believes that the first
piece is bigger than the second piece, otherwise it returns $\text{smaller}$.
Finally, $\kwalloc$ determines the final allocation by specifying which agent
gets which piece.


For this protocol, envy-freeness is easy to show. The first agent is
indifferent between the two pieces since they initially split the cake. If the
second agent believes that
$\mathit{piece}$ is bigger than $\mathit{piece}'$, then the second agent
receives $\mathit{piece}$ and has no envy; similarly, if the second agent
believes that $\mathit{piece}$ is smaller than $\mathit{piece'}$, then the
second agent receives $\mathit{piece}'$ and again has no envy. Thus, no matter
what the final allocation is, neither agent has envy: Cut-Choose always
produces an \emph{envy-free} allocation.

\subsection{Reducing Envy-Freeness to Constraint Satisfiability}

To automate proofs of envy-freeness, we define a translation of programs into
constraints in first-order logic, which we can dispatch to an SMT solver. Our
translation is compositional, guided by the structure of the program. At a
high-level, our approach has three steps.

First, we convert a program $e$ into a first-order formula $C[e]$ with one
free variable $\nu_a$ for each agent, representing agent $a$'s valuation
function, and a special free variable $\alpha$, representing the final
allocation. Then, supposing $\mathcal{A} = \{1,\ldots ,n\}$, we build a
predicate $\mathit{EnvyFree}(\nu_1, \dots, \nu_n, \alpha)$ asserting that
$\alpha$ is envy-free for valuations $\nu_1, \dots, \nu_n$.  Similarly, we build
a predicate $\mathit{ValidValn}(\nu_1, \dots, \nu_n)$ asserting that the
valuations satisfy the assumptions we introduced at the start of this section.
Finally, we combine our predicates into a single constraint capturing
envy-freeness, where $\overline{\nu}$ stands for the sequence of valuation
variables $\nu_1, \dots, \nu_n$:
\begin{equation}
  \label{form:envyfree}
  \forall \overline{\nu}.\, \forall \alpha.\, C[e](\nu, \alpha) \land
  \mathit{ValidValn}(\overline{\nu})
  \implies \mathit{EnvyFree}(\overline{\nu}, \alpha) .
\end{equation}
In words, this formula states that for all valuation functions $\nu_1, \dots,
\nu_n$ and all possible allocations $\alpha$, if $\alpha$ is a possible output
from the program $e$ for the given valuations, and the given valuations satisfy
the axioms, then the allocation $\alpha$ is envy-free. (We have elided some
details here; full details can be found in \Cref{sec:constraints}.) 



In our running example, the protocol Cut-Choose is translated to the following
constraint:
\[
  C[\text{Cut-Choose}]
  = \exists \rho, \rho'.\, \underbrace{\nu_1(\rho) = \nu_1(\rho')}_{\kwsplit_1}
  \land \text{ if } \underbrace{\nu_2(\rho) \geq \nu_2(\rho')}_{\text{compare}_2}
  \text{ then } \underbrace{\alpha = (\rho', \rho)}_{\text{allocate}}
  \text{ else } \underbrace{\alpha = (\rho, \rho')}_{\text{allocate}} .
\]
The existential variables $\rho$ and $\rho'$ correspond to the program variables
$\mathit{piece}$ and $\mathit{piece}'$, respectively.  The annotations describe
the program operations that correspond to the different subformulas.

By combining the above formula with the predicates $\mathit{ValidValn}$ and
$\mathit{EnvyFree}$ as in \eqref{form:envyfree}, we arrive at a first-order
logical formula stating that Cut-Choose is envy-free for all valuations
satisfying the axioms. This property can then be checked by an automated solver.

\section{\SYSTEM: A Language for Fair Division}
\label{sec:language}

Now that we have seen how our system works at a high level, we turn to defining
our language for fair division protocols, \SYSTEM. We first introduce the query
model, then present the syntax and operational semantics of our language.

\subsection{Robertson-Webb Query Model}
As we discussed, protocols in fair division do not have direct access to agent
valuations; instead, the protocol asks agents to perform specific actions
depending on their personal valuations. A typical way to formalize and compare
protocols with different allowed operations is through a \emph{query model},
which describes the queries about the agent valuation functions that the
protocol is allowed to make. To implement the queries in practice, the protocol
can ask agents to answer a query, or perform some action.



Much of the cake-cutting literature has converged on the \emph{Robertson-Webb}
(RW) query model, which enables rich protocols while assuming
queries that are realistic to implement in practice. Introduced by
\citet{robertson1998cake} and further refined (and named) by
\citet{WOEGINGER2007213}, this model allows two kinds of queries:
\begin{description}
  \item[Eval.] Given an interval $[\ell, r]$ and an agent $a$, return agent $a$'s
    valuation for the interval: $V_a([\ell, r])$.
  \item[Mark.] Given a left-endpoint $\ell$, an agent $a$, and a target value $v \in
      [0, 1]$ such that $v \leq V_a([\ell, 1])$, return $r$ such that $V_a([\ell,
          r]) = v$.
\end{description}
The eval query is straightforward, but the mark query is more subtle. First,
the mark query returns a point $r$ in the unit interval $[0, 1]$, intuitively,
a mark where the cake could be cut, without actually cutting
the cake at that position. The protocol is free to use the mark in future
computations, for instance comparing different marks to see which one is
largest. Second, without some assumptions on the valuation functions, it is not
clear that a mark $r$ always exists---for instance, if the valuation
\emph{jumps} from $v - 0.1$ to $v + 0.1$, it would be impossible to return a
mark describing a piece with value exactly equal to $v$. However, the
\textbf{Continuity} assumption introduced in \Cref{sec:motiv} ensures that such
jumps cannot happen.
Finally, note that that the mark $r$ may not be unique: there may be more than
one possible mark for a particular target value $v$. The protocol has no control
over how the agent selects the mark, so the mark operation is non-deterministic.

\paragraph{Example: Implementing operations in the RW model.}
Specific agent actions and operations can be implemented on top of the query
model. For instance, each of the four operations we introduced in
\cref{sec:motiv}---\textbf{Divide}, \textbf{Trim}, \textbf{Compare}, and
\textbf{Select}---can be implemented using RW queries. The operations
\textbf{Compare} and \textbf{Select}, which ask an agent to compare two pieces,
or select their favorite piece, can be implemented by making eval queries and
comparing the values.
The operations \textbf{Divide} and $\mathbf{Trim}$ can be implemented using an eval query and a mark
query. 

For instance, suppose we want to divide an interval $[\ell, r]$ into two
pieces that agent $a$ considers to be equally good. We first make an eval query
to get $a$'s value for $[\ell, r]$; call this value $z$ and suppose that it is
non-zero. Then, we make a mark query starting from $\ell$, with target value
$z/2$. Since $z/2$ is strictly less than the value of $[\ell, r]$, the mark
query will return a point $c \in [\ell, r]$ such that $[\ell, c]$ has value
exactly $z/2$. Since we assume that valuations satisfy \textbf{Additivity}
(\Cref{sec:motiv}), the remaining piece $(c, r]$ must also have value exactly $z
/ 2$. Thus, we can produce the intervals $[\ell, z]$ and $[z, r]$.

\subsection{Language Syntax}

\begin{figure}
  \begin{bnf*}
    \bnftd{e} \bnfpo
    \bnfmore{%
      n \in \NN
      \bnfor r \in \RR
      \bnfor \bnfts{$\kwtrue$}
      \bnfor \bnfts{$\kwfalse$}
      \bnfor o(e_1, \dots, e_n) \qquad (o \in \mathcal{O})
    }\\
    \bnfmore{%
      \bnfor x \in \var
      \bnfor \bnfts{$\kwlet$}\ x = e_1\ \bnfts{$\kwin$}\ e_2
      \bnfor (e_1, \ldots , e_{n})
      \bnfor \pi_{n}\ e
      \bnfor \bnfts{$\kwif$}\ e_1\ \bnfts{$\kwthen$}\ e_2\ \bnfts{$\kwelse$}\ e_3
    }\\
    \bnfmore{%
      \bnfor \bnfts{$\kwinit$}
      \bnfor \bnfts{$\kwlt$}\ e
      \bnfor \bnfts{$\kwrt$}\ e
      \bnfor \bnfts{$\ediv{e_1}{e_2}$}
    }\\
    \bnfmore{%
      \bnfor \bnfts{$\emark{a}{e_1}{e_2}$}
      \bnfor \bnfts{$\eeval{a}{e}$} \qquad (a \in \mathcal{A})
    }\\
    v \bnfpo
    \bnfmore{%
      n \in \NN
      \bnfor r \in \RR
      \bnfor \bnfts{$\kwtrue$}
      \bnfor \bnfts{$\kwfalse$}
      \bnfor (v_1, \ldots ,v_{n})
      \bnfor \left[ r_1, r_2 \right] \qquad (r_1 \leq r_2)
    }
  \end{bnf*}
  \caption{Expressions and Values in \SYSTEM}
  \label{bnf:e}
\end{figure}

Now, we are ready to present the language. \SYSTEM is a standard first-order
language, extended with a few custom constructs. We briefly walk through the
syntax of expressions and values, presented in \Cref{bnf:e}. Constants can be natural
numbers, real numbers, or booleans. We assume a fixed collection $\ops$ of
primitive operations $o$, such as the usual comparisons (e.g., equality $e_1 =
  e_2$ and lesser-than $e_1 \leq e_2$) and boolean operations, and arithmetic
operators (e.g., addition and multiplication). Variables $x$ in the
language are bound in let-bindings; we assume that variable names are drawn
from a countably infinite set $\var$. Rounding out the standard constructs, the
language has $n$-ary products and projections, and conditionals.

The remaining expressions are particular to cake-cutting protocols. First, we
have operations to manipulate intervals, which represent pieces of the cake.
The unit interval (the whole cake) is represented by $\kwinit$. The end points
of an interval can be computed using $\kwlt\ e$ and $\kwrt\ e$, respectively.
The operation $\ediv{e_1}{e_2}$ splits an interval $e_1$ into two sub-intervals
by cutting at at location $e_2$, a real number. Finally, we have constructs
from the Robertson-Webb model: $\emark{a}{e_1}{e_2}$ is a mark query on
left-endpoint $e_1$, with target value $e_2$, and $\eeval{a}{e}$ is an eval
query on interval $e$. These operations are indexed by an agent $a$; we assume
that $\agt$ is a fixed, finite set of agents. Pieces, which consist of a finite
set of intervals, are represented using tuples of intervals.
The values are entirely standard. Intervals $[r_1, r_2]$ are represented by
pairs of real numbers satisfying $r_1 \leq r_2$.  Variables cannot appear in
values.



We give our language a simple type system with base types (booleans $\BB$,
natural numbers $\NN$, real numbers $\RR$, positions $\mathbb{E}$ in $[0, 1]$,
and intervals $\II$) and $n$-ary products; we write $\typ$ for the set of types.
We elide the description of the type system, which is entirely standard, and
assume throughout that all programs are well-typed.

\subsection{Operational Semantics}

\begin{figure}
  \begin{mathpar}
    \inferrule*[right=Val]
    { }
    {\left< v, \sigma \right> \Downarrow v}
    \and
    \inferrule*[right=Ops]
    {%
    \left< e_1,\sigma \right> \Downarrow v_1 \\
    \cdots \\
    \left< e_n, \sigma \right> \Downarrow v_n \\
    \sem{o} : \val^n \to \val
    }
    {\left< o(e_1, \ldots ,e_{n}), \sigma \right> \Downarrow \sem{o}(v_1,\ldots ,v_{n})}
    \\
    \inferrule*[right=Pair]
    {%
    \left< e_1,\sigma \right> \Downarrow v_1 \\
    \cdots \\
    \left< e_n, \sigma \right> \Downarrow v_n
    }
    {\left< (e_1, \ldots ,e_{n}), \sigma \right> \Downarrow (v_1,\ldots ,v_{n})}
    \and
    \inferrule*[right=Proj]
    {%
      \left<e, \sigma \right> \Downarrow (v_1 ,\ldots ,v_{n}) \\
      1 \leq k \leq n
    }
    {\left< \pi_{k}\ e, \sigma \right> \Downarrow v_{k}}
    \\
    \inferrule*[right=IfTrue]
    {%
      \left< e_1, \sigma \right> \Downarrow \kwtrue \\
      \left< e_2, \sigma \right> \Downarrow v_2
    }
    { \left< \kwif\ e_1\ \kwthen\ e_2\ \kwelse\ e_3 , \sigma \right> \Downarrow v_2}
    \and
    \inferrule*[right=IfFalse]
    {%
      \left< e_1, \sigma \right> \Downarrow \kwfalse \\
      \left< e_3, \sigma \right> \Downarrow v_3
    }
    { \left< \kwif\ e_1\ \kwthen\ e_2\ \kwelse\ e_3 , \sigma \right> \Downarrow v_3}
    \\
    \inferrule*[right=Let]
    {%
      \left< e_1, \sigma \right> \Downarrow v_1 \\
      \left< e_2, \sigma[x \mapsto v_1]\right> \Downarrow v_2
    }
    {\left< \kwlet\ x = e_1\ \kwin\ e_2, \sigma \right> \Downarrow v_2}
    \and
    \inferrule*[right=Var]
    { }
    {\left< x, \sigma \right> \Downarrow \sigma(x)}
    \\
    \inferrule*[right=Lt]
    { \langle e, \sigma \rangle \Downarrow [r_1, r_2] }
    {\left< \kwlt\ e, \sigma \right> \Downarrow r_1}
    \and
    \inferrule*[right=Rt]
    { \langle e, \sigma \rangle \Downarrow [r_1, r_2] }
    {\left< \kwrt\ e, \sigma \right> \Downarrow r_2}
    \and
    \inferrule*[right=Cake]
    { }
    {\left< \kwinit, \sigma \right> \Downarrow \left[ 0,1 \right]}
    \\
    \inferrule*[right=Div]
    {%
      \left< e_1, \sigma \right> \Downarrow [r_1, r_1'] \\
      \left< e_2, \sigma \right> \Downarrow r_2 \\
      r_1 \leq r_2 \leq r_1'
    }
    {\left< \ediv{e_1}{e_2}, \sigma \right> \Downarrow ([r_1, r_2], [r_2, r_1'])}
    \\
    \inferrule*[right=Mark]
    {%
      \left< e_1, \sigma \right> \Downarrow v_1 \\
      \left< e_2, \sigma \right> \Downarrow v_2 \\
      V_a(\left[v_1, r\right]) = v_2
    }
    {\left< \emark{a}{e_1}{e_2}, \sigma \right> \Downarrow r}
    \and
    \inferrule*[right=Eval]
    {%
      \left< e, \sigma \right> \Downarrow [r, r']
    }
    {\left< \eeval{a}{e}, \sigma \right> \Downarrow V_a([r, r'])}
  \end{mathpar}

  \caption{Big-step Operational Semantics of \SYSTEM}
  \label{rules}
\end{figure}

We define the semantics of our programs in big-step style, with judgements of
the form:
\[
  \langle e, \sigma \rangle \Downarrow v.
\]
Above, the \emph{environment} $\sigma : \var \rightharpoonup \val$ is a partial
map assigning values to variables; we call a pair of an expression and an
environment $\langle e, \sigma \rangle$ a \emph{configuration}. We assume
variables are typed, so there are disjoint sets $\var_\tau$ for each type $\tau
  \in \typ$, and we write $\val_\tau$ for the set of values of type $\tau$. To
reduce the notation, we elide type ascriptions in our presentation of the formal
system.

Our operational semantics is parametrized by a set of valuation functions $\{
  V_a \}_a$, one for each agent $a \in \agt$, where each valuation satisfies the
conditions in \Cref{sec:motiv}. \Cref{rules} presents the rules of our
operational semantics. Most of the rules are standard. The variable rule
\rname{Var} looks up the value of a variable in the environment. For operations
\rname{Ops}, we assume that each operation $o \in \ops$ of arity $n$ is
interpreted as a function $\sem{o}$ from $\val^n$ to $\val$; operations have
types given by some signature. Pairing \rname{Pair} and projections
\rname{Proj} are standard, as are the rules for if-then-else (\rname{IfTrue}
and \rname{IfFalse}). Let-binding \rname{Let} evaluates $e_1$ to a value $v_1$,
extends the environment with a new binding mapping $x$ to $v_1$, and then
evaluates the body.

The remaining rules manipulate intervals and implement queries.  \rname{Cake}
represents the whole cake, and evaluates to the unit interval $[0, 1]$;
\rname{Lt} and \rname{Rt} get the left and right endpoints of an interval.
\rname{Div} splits an interval at a certain position into a pair of intervals;
the side condition requires that the split point must be within the bounds of
the interval. \rname{Mark} evaluates its first argument to a position in $[0,
1]$, and its second argument to a target value $v$. If there is a position $r$
such that $V_a([v_1, r])$ is exactly equal to $v$, then the program can return
$r$. Finally, \rname{Eval} simply computes the value of an interval.

\paragraph*{Non-determinism.} As we have seen, mark queries can be
non-deterministic if an agent $a$ is indifferent between $[\ell, r]$ and
$[\ell, r']$, i.e., if agent $a$ has value zero for $[r, r']$.
%
%
Thus, it is not possible to eliminate let-bindings in our language because the
following programs are not equivalent:
\[
  \kwlet\ x = e_1\ \kwin\ e_2 \;\; \nequiv \;\; e_2\{ e_1 / x \}
\]
For instance, if $e_1$ is non-deterministic and the body $e_2$ mentions $x$
twice, then the program on the left makes the non-deterministic choice once,
while the program on the right makes the non-deterministic choice twice.

\paragraph*{Errors.} A configuration $\langle e, \sigma \rangle$ where $e$ does not
step but is not a value is known as a \emph{stuck configuration}, and is the
result of a protocol error. Since we assume that programs are well-typed, there
are only two ways a program can get stuck. In \rname{Div}, a program becomes
stuck if it attempts to cut outside the bounds of the given interval. In
\rname{Mark}, a program becomes stuck if it is not able to find a point $r$ so
that $[v_1, r]$ has value $v_2$. By our assumptions on valuations, this can
only happen if target value $v_2$ is strictly greater than the value of $[v_1,
      r]$.

\paragraph*{Example: Encoding Cut-Choose.} Putting everything together, suppose
that we have two agents $\agt = \{ 1, 2 \}$. \Cref{proto:cc} shows how to
express the Cut-Choose protocol from \Cref{sec:motiv} in \SYSTEM. The output
allocation is represented as tuple: the first component is allocated to agent
$1$, while the second component is allocated to agent $2$.

\begin{figure}
  \[
    \begin{array}{l}
      \kwlet\ \mathit{pieces} = \ediv{\kwcake}{\emark{1}{0}{1/2}}\ \kwin                      \\
      \kwif\ \eeval{2}{\pi_1\ \mathit{pieces}} \geq \eeval{2}{\pi_2\ \mathit{pieces}}\ \kwthen \\
      \quad (\pi_2\ \mathit{pieces}, \pi_1\ \mathit{pieces})                                  \\
      \kwelse                                                                                 \\
      \quad (\pi_1\ \mathit{pieces}, \pi_2\ \mathit{pieces})
    \end{array}
  \]
  \caption{Cut-Choose in \SYSTEM}
  \label{proto:cc}
\end{figure}

\section{Constraint generation, soundness, and completeness}
\label{sec:constraints}

Now that we have seen the language, we turn to verifying properties about
programs.  First, we define a translation from programs $e$ to constraints $c(e,
\ret)$, where $\ret$ is a special variable. Intuitively, $c(e, v)$ should hold
exactly when $e$ can step to $v$; we formalize this claim by proving soundness
and completeness of our translation. Finally, we express envy-freeness as a
formula $E(\ret)$, and verify this property by checking if $c(e, \ret)$ implies
$E(\ret)$ for all possible outputs $\ret$.

\subsection{Constructing the constraints}

Our constraint translation targets a standard, multisorted first-order logic.

\paragraph{Syntax of formulas}
The signature $(S, C, F, Q)$ of our logic contains sorts, constants, function
symbols, and predicates respectively. The sorts in $S$ correspond exactly to the
set of types $\mathcal{T}$; we denote the sort corresponding to $\tau$ by
$S_{\tau}$ for each $\tau \in \mathcal{T}$. The set of constants $C$ consists of
$\mathbb{N}$ and $\{\mathsf{true}, \mathsf{false}\}$. The set of function
symbols is
\[
  F = \mathcal{O}\cup \{(\_,\ldots ,\_)_{k}, \pi_{k} \mid k= 1,2, \ldots \}\cup \{\ite{}{}{}, \ell, r, [ \_,\_ ]\}\cup \{\v_{a} \mid a \in \mathcal{A}\}
\]
with the expected arities (e.g., $\ite{}{}{}$ (if-then-else) has arity $3$, the
valuations $\v_a$ have arity $1$, etc.).
%
%
Finally, the set of predicate symbols $Q$ consists of $=$ and $\geq$.

We consider three disjoint sets of logical variables: $\mathcal{Y} = \{y_1,y_2,
y_3 ,\ldots \}$, $\mathcal{X}$, and $\{\ret\}$. We refer to the set of logical
terms as $\term$ and the set of formulas as $\f$. As usual, logical terms are
formed by variables, constants, and function symbols applied to other terms, and
formulas are constructed from predicates, connectives, and quantifiers from
first-order logic ($\neg$, $\wedge$, $\vee$, $\Rightarrow$, $\forall$, $\exists$).

\paragraph{Semantics of formulas}
An interpretation of our logic interprets sorts as sets, function symbols as
functions, and predicate symbols as predicates. We fix the following
interpretation of our logic.
\begin{definition}
  Let $\interp$ be an interpretation. We say that $\interp$ is \emph{proper} if
  \begin{enumerate}
    \item $\interp(S_{\tau}) = \val_{\tau}$ for all $\tau \in \mathcal{T}$;
    \item $\interp(o) = \sem{o}$ for any $o \in \mathcal{O}$;
    \item $\interp( (\_,\ldots,\_))$ and $\pi$ are the constructor and
          projection of a product respectively, and $\interp(\ite{}{}{})(a,b,c)$ is
          $b$ if $a$ is true, else $c$ if $a$ is false;
    \item $\interp(\ell)$ and $\interp(r)$ give the left and right endpoints of
          intervals respectively, and $\interp([\_, \_])$ is the interval
          $[r_1,r_2]$ given real numbers $r_1$ and $r_2$;
    \item $\interp(\v_{a})$ is a valuation function for any $a$ in any agent set $\mathcal{A}$.
  \end{enumerate}
  We let $\M$ denote the class of proper interpretations. Given an agent set
  $\mathcal{A}$ and a set of valuation functions $\vset = \{V_{a} \mid a \in
  \mathcal{A}\}$, we let $\M_{\vset}$ denote the class of proper interpretations
  where $\v_{a}$ is interpreted to be $V_{a}$ for all $a \in \mathcal{A}$.
\end{definition}

We assume that interpretations also interpret all logical variables: if $y$ is
some variable of sort $S_{\tau}$, then $\interp(y) \in \interp(S_{\tau})$. The
interpretation extends to all terms; we write $\interp(t)$ for the
interpretation of the logical term $t$. Finally, an interpretation determines
which formulas are valid. We write $\interp \vDash \varphi$ if the formula
$\varphi$ holds in the interpretation, $\vDash \varphi$ if all interpretations
in $\M$ satisfy $\varphi$, and $\M_{\vset} \vDash \varphi$ if all
interpretations in $\M_{\vset}$ satisfy $\varphi$.

\paragraph{From expressions to formulas}
We will translate programs from our core language into logical formulas. To
separate the syntax of the programming language from the syntax of our logical
formulas, we will use a hat to denote the logical counterpart of some program
syntax. For instance, we write $\vemb{v}$ for the logical term corresponding to
the program value $v\in \val$, and $\vemb{x}$ for the logical variable
corresponding to the program variable $x \in \mathcal{X}$.

We will define a constraint translation with the following type:
\[
  \emap : \mathbb{N}\times \exp \to \mathbb{N} \times \f \times \term,
\]
and we let $\emap_{Q}$, $\emap_{s}$, and $\emap_{\rho}$ be the first, second,
and third projections of $\emap$ respectively. Intuitively, a program
expression $e$ is translated into a logical term $\emap_{\rho}$ which encodes
the values $v$ that $e$ can step to. There are two complications with modeling
the evaluation relation. First, some rules (like \rname{Div} and \rname{Mark})
have side-conditions that must hold; the formula $\emap_{s}$ tracks these
side-conditions. Second, since the evaluation relation is non-deterministic,
a single expression may step to multiple different values. Our constraint
translation models non-deterministic choice by introducing new logical
variables; the natural-number input and output $\emap_{Q}$ track the number of
logical variables to ensure that new logical variables are fresh.

\Cref{qsr:a} gives the definition of $\emap$ for the standard constructs. We let
$\n{e} = \emap_Q(0, e)$ and $\n{e_1 ,\ldots ,e_{n}} = \n{e_1} + \cdots +
\n{e_{n}}$; intuitively, these quantities count the number of fresh logical
variables that are introduced when translating the expressions.  Most of the
cases are straightforward; we comment on the more interesting aspects. For
operators, $o$ ranges over the primitive operations $\mathcal{O}$, as well as
constructs like $\kwlt, \kwrt$, pairing and projection. For if-then-else, the
side condition $\emap_{s}$ conjoins the side-condition of the guard $e_1$ with a
case analysis: if the guard is true $(\emap_{\rho}(k, e_1) = \kwtrue)$ then the
side-condition for the first branch $e_2$ should hold; otherwise, the
side-condition for the second branch $e_3$ should hold. The logical term
$\emap_{\rho}$ is then just if-then-else applied to the logical terms for the
guard and the branches.

\Cref{qsr:b} gives our constraint translation for the more specialized
constructs in our language. In the translation for $\kwdiv$ and $\kwmark$, the
last two conjuncts in the side-conditions reflect the side-conditions in the
operational rules \rname{Div} and \rname{Mark}. Finally, the translation for
$\kwmark$ introduces a new logical variable $y$, with index chosen so that it is
guaranteed to be fresh. This program construct is the source of non-determinism
in the language.


Finally, we define the \emph{constraint of $e$}:
\[
  c(k, e, t) \triangleq \emap_{s}(k,e) \wedge (t = \emap_{\rho}(k,e)).
\]
We will write $c(e,t) \triangleq c(0,e,t)$ for short. If we close this formula by
existentially quantifying over all logical variables, this formula states that
the logical term $t$ describes a value that $e$ can step to. For instance,
$\exists y_1, \dots, y_{N(e)}.\, c(e, \vemb{v})$ states that $e$ can reduce to
$v$.
\paragraph{Example} Translating the program Cut-Choose from \Cref{proto:cc} and
simplifying, we have:
\begin{equation*}
  \begin{array}{rl}
    \emap_Q(0, e)  &=    1                                                                                                                        \\
    \emap_{s}(0, e) &=     r([0,1]) \geq y_1\geq \ell([0,1])                                                                                     \\
                        &\quad \wedge (\v_1 (\left[ 0,y_1 \right]) = 1/2) \wedge (\v_1([0, 1])\geq y_1)                                                 \\
                        &\quad \wedge ((\v_2 (\pi_1 (\alpha) ) \geq \v_2 (\pi_2 (\alpha))) = \mathsf{true})
                        \vee ((\v_2 (\pi_1 (\alpha) ) \geq \v_2 (\pi_2 (\alpha))) = \mathsf{false}))\\
    \emap_{\rho}(0, e) &=  \ite{\v_2 (\pi_1 (\alpha)) \geq \v_2 (\pi_1 (\alpha))}{(\pi_2 (\alpha), \pi_1(\alpha))}{(\pi_1 (\alpha),\pi_2 (\alpha))}
  \end{array}
\end{equation*}
where $\alpha$ is shorthand for $( [\ell( [0,1]), y_1], [y_1, r([0,1])])$,
representing the two pieces of the cake after splitting at $y_1$. Note that
while the original expression contained variables $\mathit{cake}$ and
$\mathit{pieces}$, those were substituted away by $[0,1]$ and $\alpha$
respectively when translating the let-bindings.

We step through the three components. First, because there is one mark query in
this example, the constraint translation introduces one logical variable $y_1$
and we have $\n{e} = \emap_Q(0,e) = 1$. Second, the side-condition $\emap_s(0,
  e)$ contains three conjuncts: the first comes from the divide expression, the
second from the mark expression, and the third from the conditional. Finally,
the term $\emap_{\rho}(0,e)$ represents the possible allocations produced by
the program.
\begin{figure}[h]
  \begin{align*}
    \emap_Q(k, v)             & \d= k & \emap_{s}(k, v)             & \d= \mathsf{true} & \emap_{\rho}(k, v)             & \d= \emb{v}     \\
    \emap_Q(k, x)             & \d= k & \emap_{s}(k, x)             & \d= \mathsf{true} & \emap_{\rho}(k, x)             & \d= \vemb{x}    \\
    \emap_Q(k, \mathsf{cake}) & \d= k & \emap_{s}(k, \mathsf{cake}) & \d= \mathsf{true} & \emap_{\rho}(k, \mathsf{cake}) & \d= \emb{[0,1]}
  \end{align*}
  \begin{align*}
    \emap_Q(k, o(e_1 , \ldots , e_n))                       & \d= k +\n{e_1 ,\ldots ,e_{n}}                                                                                             \\
    \emap_{s}(k, o(e_1 , \ldots , e_n))                     & \d= \emap_s(k, e_1) \wedge \emap_s(k + \n{e_1}, e_2) \wedge \cdots \wedge \emap_s(k + \n{e_1 ,\ldots ,e_{n-1}}, e_{n}) \\
    \emap_{\rho}(k, o(e_1 , \ldots , e_n))                  & \d= \widehat{o}(\emap_{\rho}(k, e_1), \ldots , \emap_{\rho}(k + \n{e_1 ,\ldots ,e_{n-1}}, e_n) )                       \\ \\
    \emap_Q(k, \kwif\ e_1\ \kwthen\ e_2\ \kwelse\ e_3)      & \d= k + \n{e_1, e_2 ,e_3}                                                                                              \\
    \emap_{s}(k, \kwif\ e_1\ \kwthen\ e_2\ \kwelse\ e_3)    & \d= \emap_{s}(k, e_1)\\
                                                            & \wedge ((\emap_{\rho}(k, e_1) = \mathsf{true})\wedge \emap_{s}(k + \n{e_1}, e_2))                                      \\
                                                            & \quad \vee ((\emap_{\rho}(k, e_1) = \mathsf{false})\wedge \emap_{s}(k + \n{e_1,e_2}, e_3)))                            \\
    \emap_{\rho} (k,\kwif\ e_1\ \kwthen\ e_2\ \kwelse\ e_3) & \d=  \ite{\emap_{\rho}(k, e_1)}{\emap_{\rho}(k + \n{e_1}, e_2)}{\emap_{\rho}(k + \n{e_1,e_2}, e_3)}                    \\ \\
    \emap_Q(k, \kwlet\ x = e_1\ \kwin\ e_2)                 & \d= k + \n{e_1, e_2}                                                                                                   \\
    \emap_{s}(k, \kwlet\ x = e_1\ \kwin\ e_2)               & \d= \emap_s(k,e_1)\wedge \emap_s(k + \n{e_1}, e_2)\{\emap_{\rho}(k,e_1)/x\},                                           \\
    \emap_{\rho}(k, \kwlet\ x = e_1\ \kwin\ e_2)            & \d= \emap_{\rho}(k + \n{e_1}, e_2)\{\emap_{\rho}(k, e_1)/x\}                                                           \\ \\
  \end{align*}
  \caption{Constraint translation for basic language constructs.}
  \label{qsr:a}
\end{figure}
\begin{figure}[h]
  \begin{align*}
    \emap_Q(k, \ediv{e_1}{e_2})          & \d= k + \n{e_1,e_2}                                                                                                  \\
    \emap_{s}(k,  \ediv{e_1}{e_2})       & \d= \emap_{s}(k, e_1)\wedge \emap_{s}(k + \n{e_1}, e_2)                                                              \\
                                         & \quad \wedge (\emap_{\rho}(k +
      \n{e_1}, e_2) \geq
    \ell(\emap_{\rho}(k, e_1)))                                                                                                                                 \\
                                         & \quad \wedge (r(\emap_{\rho}(k, e_1)) \geq \emap_{\rho}(k + n(e_1), e_2))                                            \\
    \emap_{\rho}(k, \ediv{e_1}{e_2})     & \d=
    (\left[\ell(\emap_{\rho}(k, e_1)), \emap_{\rho}(k + \n{e_1}, e_2)  \right],
    \left[\emap_{\rho}(k + \n{e_1}, e_2), r(\emap_{\rho}(k, e_1))  \right])                                                                                     \\ \\
    \emap_Q(k, \emark{a}{e_1}{e_2})      & \d= k + \n{e_1,e_2} + 1                                                                                              \\
    \emap_{s}(k, \emark{a}{e_1}{e_2})    & \d= \emap_{s}(k, e_1)\wedge \emap_{s}(k + \n{e_1}, e_2)                                                              \\
                                         & \quad \wedge (\v_{a}(\left[ \emap_{\rho}(k, e_1), y_{k + \n{e_1,e_2} + 1} \right]) = \emap_{\rho}(k + \n{e_1}, e_2)) \\
                                         & \quad \wedge (\v_{a} (\left[ \emap_{\rho}(k, e_1), 1 \right])\geq \emap_{\rho}(k + \n{e_1}, e_2))                    \\
    \emap_{\rho}(k, \emark{a}{e_1}{e_2}) & \d= y_{k + \n{e_1,e_2} + 1}                                                                                          \\ \\
    \emap_Q(k, \eeval{a}{e})             & \d= k + \n{e}                                                                                                        \\
    \emap_{s}(k,  \eeval{a}{e})          & \d= \emap_{s}(k, e)                                                                                                  \\
    \emap_{\rho}(k, \eeval{a}{e})        & \d= \v_{a}(\emap_{\rho}(k, e))
  \end{align*}
  \caption{Constraint translation for cake cutting operations.}
  \label{qsr:b}
\end{figure}
\subsection{Soundness}
When the expression $e$ is closed, the initial environment $\sigma$ does not
affect the big-step semantics of $e$ so the constraint translation only needs
to model the program $e$. However, since our language can introduce local
variables via let-binding---which cannot be eliminated since our semantics is
non-deterministic---a compositional constraint translation must also handle
open expressions. To state (and prove) a soundness theorem, we convert program
environments $\sigma$ into logical substitutions; formalizing this bridge is
the main challenge in establishing soundness.

\paragraph{Environments}
For this section, it will be useful to represent environments as a sequence of
assignments. Let $x_1,\ldots ,x_{n} \in \mathcal{X}$ be a list of program
variables, not necessarily unique.

\begin{definition}
  We say that $\sigma$ is an \emph{environment on $x_1,\ldots ,x_{n}$} if
  $\sigma$ is a finite sequence $[x_1 \mapsto v_1][x_2 \mapsto v_2]\cdots
    [x_{n} \mapsto v_{n}]$ for some $v_1,\ldots ,v_{n}\in \val$.
  The \emph{domain} of $\sigma$ is $\dom(\sigma) = \{x_1,\ldots ,x_{n}\}$
\end{definition}
An environment determines a partial map, that is, $\sigma : \mathcal{X}
  \rightharpoonup\val$ where $\sigma(x) =  v_{i}$ if $x = x_{i}$ and $i$ is the
largest $i$ such that $x = x_i$, and otherwise not defined.

\paragraph{Substitutions}
Environments map program variables to values. On the logic side, we can reflect
this data as substitutions of logical terms for logical variables. Let
$x_1,\ldots ,x_{n} \in \mathcal{X}$, again not necessarily unique.
\begin{definition}
  We say that $S$ is a \emph{substitution on $x_1,\ldots ,x_{n}$} if $S = \{t_n/\vemb{x_n}\}
    \cdots \{t_{1}/\vemb{x_{1}}\}$ for $t_1,\ldots ,t_{n}\in \term$ such that
  $\fv(t_{i}) \cap \fv(t_{j}) \subseteq \mathcal{X}$ if $i \neq j$, and
  $\fv(t_{i})\subseteq \{x_1,\ldots ,x_{i-1}\}$ for $i = 1,\ldots ,n$.

  The \emph{domain} of $S$ is $\dom(S) = \{x_1,\ldots ,x_{n}\}$. We let $m(S)$ be
  the largest natural number $k$ such that $y_{k}$ is a free variable in one of
  $t_1,\ldots ,t_{n}$.
\end{definition}
Substitutions can be applied to logical formulas and terms. Recalling the usual
notation for substitution, we write $\varphi S$ for the formula obtained by
applying $S$ to $\varphi$, and $tS$ for the term obtained by applying $S$ to
$t$. Substitution is defined syntactically in the usual way; a list of
substitutions is applied sequentially from left to right.

We can connect substitutions with environments through an interpretation.

\begin{definition}
  Let $\interp$ be a proper interpretation. We say that $\interp$ \emph{marries
    $\sigma$ and $S$} if $\interp(\vemb{x}S) = \sigma(x)$ for all $x\in \dom(\sigma) = \dom(S)$.
\end{definition}

To state soundness of our translation, we need two pieces of notation. First,
we say that two interpretations $\interp$ and $\interp'$ \emph{agree up to} $k$
if they agree on $\{ y_1, \dots, y_k \}$. Second, if $\vset = \{ V_a \mid a \in
  \agt \}$ is a set of valuation functions, then we write $\langle e, \sigma
  \rangle \Downarrow_{\vset} v$ if the configuration $\langle e, \sigma \rangle$
can reduce to $v$ when agents have valuations $\vset$.

\begin{restatable}[Soundness]{theorem}{soundness}
  \label{prop:soundness}
  Let $e$ be a well-typed expression, $\sigma$ an environment such that
  $\fv(e) \subseteq \dom(\sigma)$, $S$ a substitution, $\vset$ a set of
  valuation functions, $v\in\val$, and $\interp$ an interpretation from
  $\M_{\vset}$ that marries $\sigma$ and $S$. If $\left< e,\sigma
    \right>\Downarrow_{\vset} v$, then for any $k \geq m(S)$, there is an
  interpretation $\interp'$ from $\M_{\vset}$ that agrees with $\interp$ on $\{
    y_1, \dots, y_k \}$ such that $\interp' \vDash c(k,e,\emb{v})S$.
\end{restatable}

For the proof of \Cref{prop:soundness}, see \append{app:proofs}.
We can conclude soundness for closed expressions.

\begin{corollary}
  \label{cor:sound}
  Let $e$ be a closed expression. Let $\vset$ be a set of valuations and let
  $v$ be a value. If $\left< e,\varepsilon \right>\Downarrow_{\vset} v$, then
  $\M_{\vset} \vDash \exists y_1,\ldots ,y_{\n{e}}.c(e, \emb{v})$.
\end{corollary}

\subsection{Completeness}

Soundness states that if program $e$ can reduce to $v$, then our constraint is
satisfiable when we plug in $v$. The converse also holds.

\begin{restatable}[Completeness]{theorem}{completeness}
  \label{thm:compl}
  Let $e$ be a well-typed expression, $S$ a substitution such that
  $\fv(e)\subseteq \dom(S)$, $\vset$ a set of valuation functions, $v\in \val$,
  and $\sigma$ an environment. 
  If there exists $k \geq 0$ and interpretation $\mu$ from $\M_{\vset}$ that marries $\sigma$ and $S$ for which $\interp \vDash c(k, e, \emb{v})S$, then $\left< e,\sigma \right>\Downarrow_{\vset}v$.
\end{restatable}

For the proof of \Cref{thm:compl}, see \append{app:proofs}. We can conclude
completeness for closed expressions as a special case.

\begin{corollary}
  \label{cor:complete}
  Let $e$ be a well-typed expression with no free variables, $\vset$ be a set
  of valuation functions, and $v \in \val$. If $\M_{\vset} \vDash \exists
    y_1,\ldots ,y_{\n{e}}.c(e,\emb{v})$, then $\left< e,\varepsilon
    \right>\Downarrow_{\vset} v$.
\end{corollary}

As a result, $\left< e,\varepsilon \right>\Downarrow_{\vset} v$ if and only if
there is $\interp \in \M_{\vset}$ such that $\interp \vDash c(e, \emb{v})$, or
equivalently, $\M_{\vset} \vDash \exists y_1,\ldots ,y_{\n{e}}.c(e,\emb{v})$.
Thus, the constraint of $e$ completely characterizes its outputs.


\subsection{Verifying Envy-Freeness}

With the constraint translation set, we are able to construct constraints
ensuring properties like envy-freeness.  Let $e$ be an expression with no free
variables and $\varphi(\ret)$ a formula containing $\ret$ as a free variable,
where the sort of $\ret$ agrees with the type of $e$. We write $e\vDash
\varphi(\ret)$ if for all sets of valuations $\vset$ and values $v\in \val$, we
have
\[
  \left< e,\varepsilon \right>\Downarrow_{\vset} v\ \text{implies}\ \M_{\vset}\vDash \varphi(\emb{v}),
\]
where $\varphi(\emb{v})= \varphi(\ret)\{\emb{v}/\ret\}$.

Now, let $\v^{k}_{a}(p) \d= \v_{a}(\pi_1 p) + \cdots + \v_{a} (\pi_{k} p)$ where $t$ has sort $S_{\mathbb{E}}^{k}$, a formula representing the valuation of a piece $p$, consisting of $k$ intervals, according to agent $a$.
Suppose that $\ret$ represents the allocation computed by the program, and
$k_{a}$ is the number of intervals allocated to agent $a$. Let 
\[
  E_{a}(\ret) \triangleq
  (\v^{k_a}_{a}(\pi_{a}\ret) \geq \v^{k_{1}}_{a}(\pi_{1}\ret)) \wedge \cdots \wedge
  (\v^{k_{a}}_{a}(\pi_{a}\ret) \geq \v^{k_n}_{a}(\pi_{n}\ret)),
\]
Then, the following formula states that $\ret$ is an envy-free allocation:
\[
  E(\ret) \triangleq \bigwedge_{a \in \mathcal{A}}E_{a}(\ret).
\]
It is not hard to see that if an allocation $v$ satisfies $E(\emb{v})$ for
agent valuations $\vset$, then the allocation is envy-free with respect to
$\vset$. Thus $e$ is envy-free if $e \vDash E(\ret)$.

Now, we can leverage soundness and completeness. First, we consider the formula
\[
  \psi \triangleq \forall \ret.(\exists y_1,\ldots ,y_{\n{e}}.c(e,\ret))
  \Rightarrow E(\ret).
\]
Now for any set of valuations $\vset$ and any $v \in \val$ such that $\left<
  e,\varepsilon \right> \Downarrow_{\vset} v$, by \Cref{cor:sound}, there exists
an interpretation in $\M_{\vset}$ that satisfies $c(e,\emb{v})$, so $\M_{\vset}
  \vDash \exists y_1,\ldots ,y_{\n{e}}. c(e, \emb{v})$. Thus if $\vDash \psi$,
then $\M_{\vset}\vDash E(\emb{v})$. If $\not \vDash \psi$, then
there is some valuations $\vset$, interpretation $\interp \in \M_{\vset}$, and value
$v$ for which
\[
  \interp \vDash \exists y_1,\ldots ,y_{\n{e}}.c(e, \emb{v})\wedge \neg E(\emb{v}).
\]
By \Cref{thm:compl}, $\left< e,\varepsilon \right>\Downarrow_{\vset} v$ and so
$e \not \vDash E(\ret)$. Thus, $e \vDash E(\ret)$ if and only if $\vDash \psi$.
As we will discuss in \Cref{sec:impl}, we can use SMT solvers to check $\vDash
\psi$.

\section{Fair-Division Protocols in \SYSTEM}
\label{sec:examples}

To evaluate \SYSTEM, we have implemented several protocols from the
fair-division literature.

\subsection{Notations, Conventions, and Syntactic Sugar.}

Fair-division protocols often use a large number of branches, operations,
and temporary variables. To make our example programs easier to read, we
adopt several conventions.

\paragraph*{Metavariables and notation.} We use capital letters (usually $I$, or
$A, B, \dots$) for variables ranging over intervals representing pieces of the
cake. We use $v$ for variables ranging over values, $m$ for variables ranging
over marks (i.e., positions on the unit interval $[0, 1]$), and $a$ for
variables ranging over agents. Programs use numbers ($1, 2, 3, \dots$) to
specify agents, and we sometimes to refer to agents as P1, P2, P3, etc. for
short. To reduce the number of variables, our programs frequently
shadow/re-bind variables. Just like in \Cref{sec:language}, the subscripts on
operations like $\kwmark_a$ and $\kweval_a$ indicate which agent's valuation
should be used.

\paragraph*{Syntactic sugar.} We make use of standard syntactic sugar, e.g., for
let-binding a pair:
\[
  \kwlet\ (I, J) = e\ \kwin\ e'
  \;\triangleq\;
  \kwlet\ H = e\ \kwin\ \kwlet\ I = \pi_1\ H\ \kwin\ \kwlet\ J = \pi_2\ H\ \kwin\ e'
  \qquad \text{($H$ fresh)} .
\]
Note that we must first let-bind the expression $e$ before projecting, so that
non-deterministic choices are made just once for both components. We use the
keyword $\kwalloc$ to indicate the final allocation. For instance, $\kwalloc(A,
B)$ represents the allocation that assigns $A$ to agent $1$ and $B$ to agent
$2$. This keyword is not present in our core language (\Cref{sec:language})---it
is merely an annotation to help highlight the final allocation.

\paragraph*{User-defined abbreviations.} To reduce code repetition, we will make
use of basic user-defined abbreviations/macros. For instance, the following
abbreviation takes two pieces $I_1$ and $I_2$, and orders them in decreasing
order according to agent $a$'s valuation:
\[
  \begin{array}{l}
    \kwdef\ \kwsort_a(I_1, I_2) =              \\
    \quad\kwlet\ v_1 = \eeval{a}{ I_1 }\ \kwin \\
    \quad\kwlet\ v_2 = \eeval{a}{I_2}\ \kwin   \\
    \quad\kwif\ v_1 \geq v_2\ \kwthen\ (I_1, I_2)\ \kwelse\ (I_2, I_1)
  \end{array}
\]
We have similar, $n$-ary versions of this operation for ordering $n$ pieces.

\subsection{Cut-Choose}
\begin{figure}[h]
  \centering
  \captionsetup{justification=centering}
  \begin{minipage}[b][5cm][b]{0.49\linewidth}
    \[
      \begin{array}{l}
        \kw{CutChoose} :                                   \\
        \quad\kwlet\ m = \emark{1}{0}{1/2}\ \kwin          \\
        \quad\kwlet\ (I_1, I_2) = \ediv{\kwcake}{m}\ \kwin \\
        \quad\kwlet\ (A, B) = \kwsort_2(I_1, I_2)\ \kwin   \\
        \quad\kwalloc (B, A)
      \end{array}
    \]
    \caption{Cut-Choose in \SYSTEM.}
    \label{code:cc}

  \end{minipage}\hfill
  \begin{minipage}[b][5cm][b]{0.49\linewidth}
    \includegraphics[width = \linewidth]{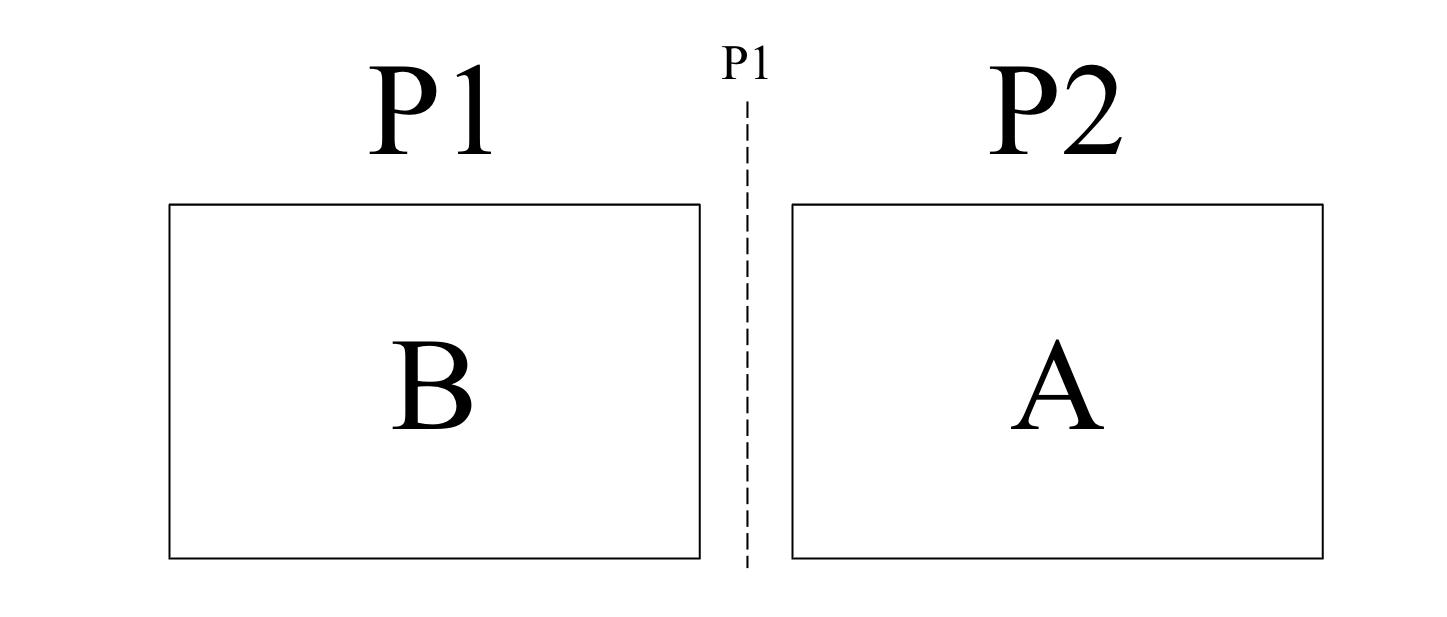}
    \caption{Possible allocation from Cut-Choose.}
    \label{fig:cc}
  \end{minipage}
\end{figure}

To warm up, let's revisit the Cut-Choose protocol from \Cref{sec:language}. We
reproduce the code from \Cref{proto:cc} in \Cref{code:cc} for convenience, now
using the abbreviation $\kwsort_2$ to compare the pieces. The code implements
the following protocol:
\begin{enumerate}
  \item P1 marks a position $m$ in $[0, 1]$ where they view the pieces to the left and
        to the right of the mark as equally good (value $1/2$).
  \item P1 cuts the cake at $m$ into two pieces, $I_1$ and $I_2$.
  \item P2 orders the pieces according to their valuation, with $A$ being their
        favorite piece and $B$ being the other piece.
  \item P1 and P2 receive $B$ and $A$, respectively.
\end{enumerate}
We can visualize this protocol using the diagram in \Cref{fig:cc}. P1 makes the
first cut, then the two pieces labeled $B$ and $A$ are allocated to the two
agents. The diagram does not indicate that P2 orders the pieces, and it only
represents one possible allocation---depending on their valuations, P2 may
prefer the piece on the left instead of the piece on the right. But the picture
serves as a rough guide for which agent cuts, and which agent receives which
piece. While Cut-Choose is simple enough to explain in words, these diagrams
will help to visualize the more complex protocols to come.

\subsection{Surplus}
In some scenarios, such as when there is a section of the cake that brings
negative utility to both agents or when there is a social benefit to conserving
a resource, it can be desirable to leave part of the cake unallocated.
Protocols that don't need to allocate the entire cake are said to work in the
\emph{free disposal} model. Of course, the protocol should try to allocate as
much of the cake as possible---a protocol that leaves the entire cake
unallocated is not very useful.

\begin{figure}[h]
  \centering
  \captionsetup{justification=centering}
  \begin{minipage}[b][5cm][b]{0.49\linewidth}
    \[
      \begin{array}{l}
        \kw{Surplus} :                                    \\
        \quad\kwlet\ m_1 = \emark{1}{0}{1/2}\ \kwin       \\
        \quad\kwlet\ m_2 = \emark{2}{0}{1/2}\ \kwin       \\
        \quad\kwlet\ (A, S) =  \ediv{\kwcake}{m_1}\ \kwin \\
        \quad\kwlet\ (T, B) =  \ediv{\kwcake}{m_2}\ \kwin \\
        \quad\kwif\ m_1 \geq m_2\ \kwthen                 \\
        \quad\quad \kwalloc(S, T)                         \\
        \quad\kwelse                                      \\
        \quad\quad \kwalloc(A, B)
      \end{array}
    \]
    \caption{Surplus in \SYSTEM.}
    \label{code:surplus}

  \end{minipage}\hfill
  \begin{minipage}[b][5cm][b]{0.49\linewidth}
    \includegraphics[width=\linewidth]{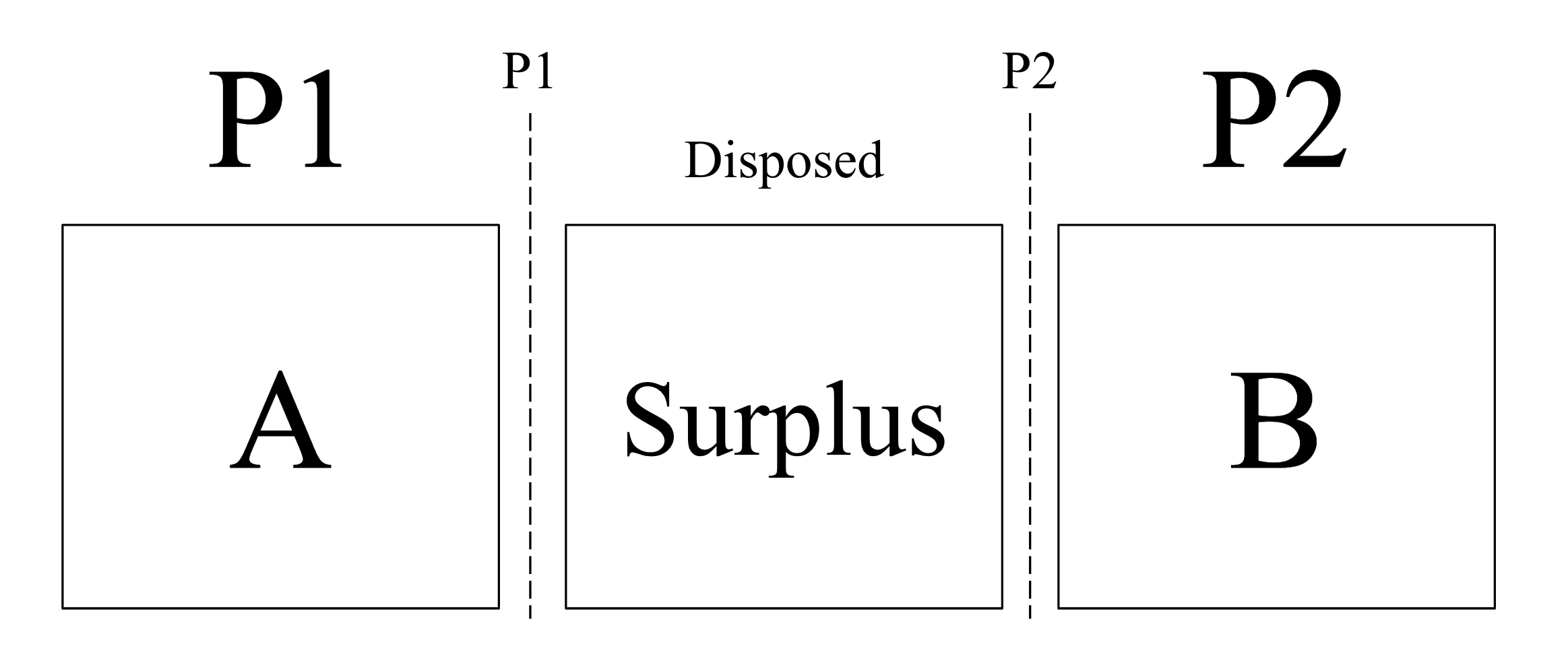}
    \caption{Possible allocation from Surplus.}
    \label{fig:surplus}
  \end{minipage}
\end{figure}

The Surplus protocol~\citep{brams2006better} divides the cake such that each of
two agents receives a connected piece that they believe is worth at least half
of the cake's total value, while possibly leaving an unallocated ``surplus''
piece. The procedure works as follows:
\begin{enumerate}
  \item Each agent marks a position such that they believe both pieces are equally
        preferred; since the two agents can have different valuations, the agents can
        mark at different positions.
  \item The agents compare marks. The agent that marked farthest to the left is
        allocated the piece to the left of their mark. The other agent is
        allocated the piece to the right of their mark.
  \item The ``surplus'' piece in between the marks is disposed (not allocated).
\end{enumerate}
\Cref{code:surplus} presents our implementation of Surplus, and
\Cref{fig:surplus} shows one possible allocation. One notable feature is that
this protocol compares marks to figure out which one is the left-most
position---this operation goes beyond the simple primitives in \Cref{sec:motiv},
requiring the queries from the Robertson-Webb model (\Cref{sec:language}).


\subsection{Selfridge-Conway}
The protocols we have seen work for $n = 2$ agents. Moving to $n = 3$ agents
makes the problem significantly more challenging. The first fair-division
protocol for three agents was proposed by Selfridge and independently by
Conway~\citep{robertson1998cake}; this protocol also serves as a building-block
for more advanced protocols. At a high level, the protocol proceeds as follows:
\begin{enumerate}
  \item P1 splits the cake into three pieces that they think are equal (value exactly
        $1/3$).
  \item P2 divides their favorite piece of the three into pieces $\mathit{Trim}$ and
        $\mathit{Rest}$ so that $\mathit{Trim}$ is equal to their second favorite
        piece.
  \item The trimmed piece $\mathit{Trim}$ and the other two non-trimmed pieces are
        chosen by the agents following the order P3, P2, P1, subject to the constraint
        that if P3 doesn't choose $\mathit{Trim}$, then P2 must choose $\mathit{Trim}$.
  \item If P2 made a trivial cut ($\mathit{Rest}$ is empty), the protocol ends here.
\end{enumerate}
Next, the extra piece $\mathit{Rest}$ is divided. For this step, it is important
to distinguish whether P2 or P3 received $\mathit{Trim}$. From P2 and P3, let PA
be the agent that received $\mathit{Trim}$, and PB be the other agent.
\begin{enumerate}[resume]
  \item PB cuts $\mathit{Rest}$ into three pieces that they think are equal.
  \item The three pieces are chosen by the agents in the order PA, P1, PB.
\end{enumerate}
\Cref{proto:sc} shows the implementation of this protocol, along with one
possible allocation. To make the program more readable, we define an
abbreviation $\kw{allocRest}$ that allocates the piece $\mathit{Rest}$; note that it
takes the two agents PA and PB as parameters, along with the remaining piece to
allocate.

Selfridge-Conway can also be modified into a free-disposal protocol by simply
discarding the piece $\mathit{Rest}$ instead of dividing it.  We call this
variant Selfridge-Conway-Surplus; details can be found in \append{app:sc-plus}.

\subsection{Waste-Makes-Haste}

\begin{figure}[p]
  \centering
  \begin{minipage}[t]{0.49\linewidth}
    \[
      \small\begin{array}{l}
        \kw{SelfridgeConway} :                                                                 \\
        \quad\kwlet\ (I_1, I_1') = \ediv{\kwcake}{\emark{1}{0}{1 / 3}}\ \kwin\                 \\
        \quad\kwlet\ (I_2, I_3) = \ediv{I_1'}{\emark{1}{0}{2 / 3}}\ \kwin\                     \\
        \quad\kwlet\ (A, B, C) = \kwsort_2(I_1, I_2, I_3)\ \kwin\                              \\
        \quad\kwif\ \eeval{2}{A}\ = \eeval{2}{B}\ \kwthen\                                     \\
        \quad\quad \kwlet\ (A, B, C) = \kwsort_3(A, B, C)\ \kwin\                              \\
        \quad\quad \kwlet\ (B, C) = \kwsort_2(B, C)\ \kwin\                                    \\
        \quad\quad \kwalloc(C, B, A)                                                           \\
        \quad\kwelse\                                                                          \\
        \quad\quad \kwlet\ m = \emark{2}{\kwlt\ A}{\eeval{2}{B}}\	\kwin\                       \\
        \quad\quad \kwlet\ (\mathit{Trim}, \mathit{Rest}) = \ediv{A}{m}\	\kwin\                \\
        \quad\quad\kwlet\ (A, B, C) = \kwsort_3(\mathit{Trim}, B, C)\ \kwin\                   \\
        \quad\quad\kwif\ A \neq \mathit{Trim}\ \kwthen\                                        \\
        \quad\quad\quad\kwlet\ (R_1, R_2, R_3) = \kw{allocRest} (2, 3, \mathit{Rest}) \ \kwin\ \\
        \quad\quad\quad\kwif\ B = \mathit{Trim}\ \kwthen\                                      \\
        \quad\quad\quad\quad \kwalloc((C, R_1), (B, R_2), (A, R_3))                            \\
        \quad\quad\quad \kwelse\                                                               \\
        \quad\quad\quad\quad \kwalloc((B, R_1), (C, R_2), (A, R_3))                            \\
        \quad\quad \kwelse\                                                                    \\
        \quad\quad\quad \kwlet\ (B, C) = \kwsort_2(B, C)\ \kwin\                               \\
        \quad\quad\quad\kwlet\ (R_1, R_2, R_3) = \kw{allocRest} (3, 2, \mathit{Rest}) \ \kwin\ \\
        \quad\quad\quad \kwalloc((C, R_1), (B, R_3), (A, R_2))                                 \\
        \quad                                                                                  \\ \\
        \quad\kwdef\ \kw{allocRest}(PA, PB, \mathit{Rest}) =                                   \\
        \quad\quad\kwlet\ v = \eeval{\text{PB}} { \mathit{Rest} } \ \kwin\                     \\
        \quad\quad\kwlet\ m_1 = \emark{\text{PB}}{\kwlt\ \mathit{Rest}}{v / 3 }\	\kwin\        \\
        \quad\quad\kwlet\ (I_1, I_1') = \ediv{\mathit{Rest}}{m_1} \ \kwin\                     \\
        \quad\quad\kwlet\ m_2 = \emark{\text{PB}}{\kwlt\ I_1'}{(2/3) * v }\	\kwin\             \\
        \quad\quad\kwlet\ (I_2, I_3) = \ediv{I_1'}{m_2} \ \kwin\                               \\
        \quad\quad\kwlet\ (I_1, I_2, I_3) = \kwsort_\text{PA} (I_1, I_2, I_3) \ \kwin\         \\
        \quad\quad\kwlet\ (I_2, I_3) = \kwsort_1 (I_2, I_3) \ \kwin\                           \\
        \quad\quad\kwalloc(I_2, I_1, I_3)
      \end{array}\normalsize
    \]

    \vfill

    \includegraphics[width=\linewidth, height=4cm]{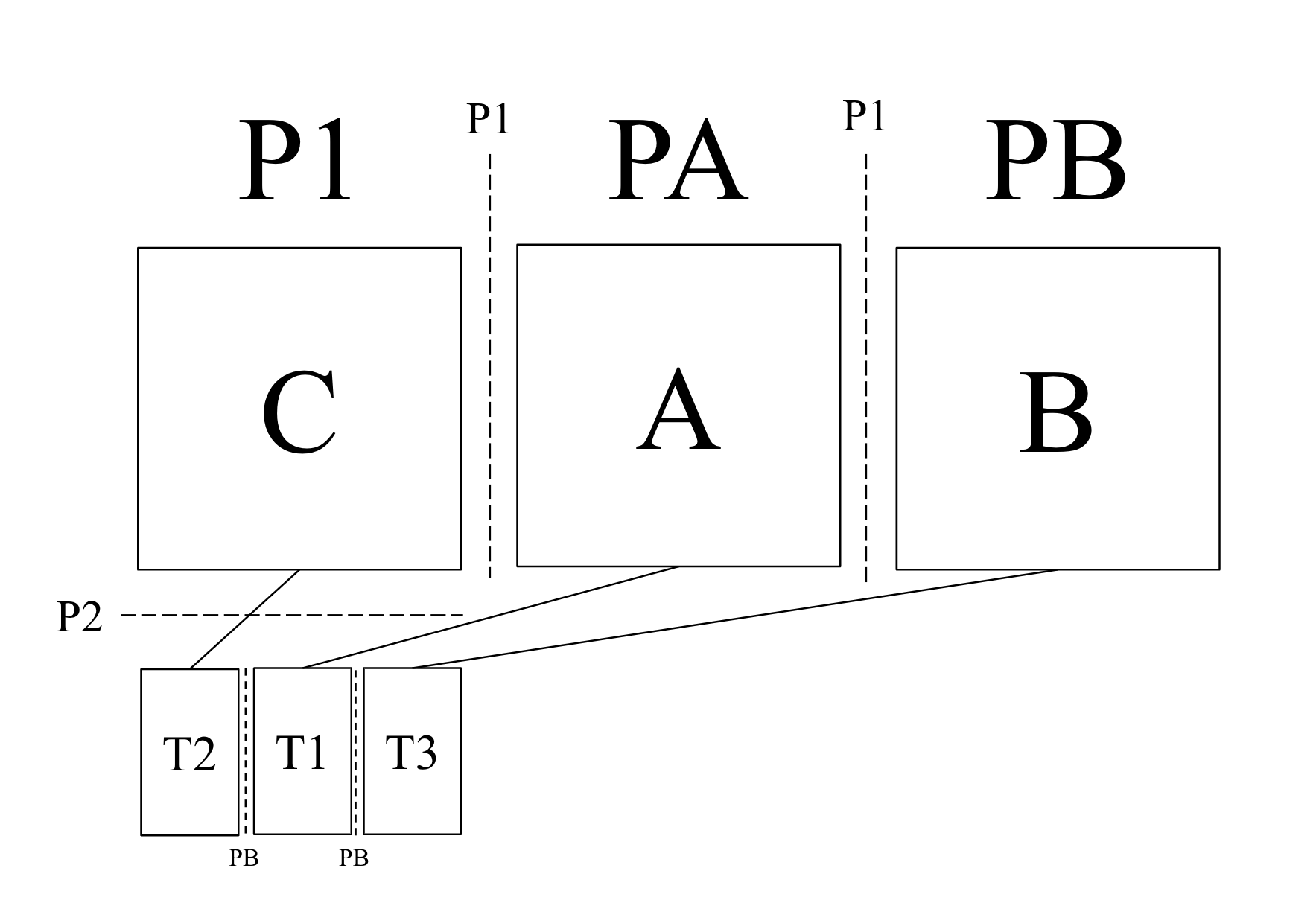}
    \begin{center}\text{Possible allocation from Selfridge-Conway}\end{center}

    \caption{Selfridge-Conway in \SYSTEM}
    \label{proto:sc}
  \end{minipage}
  \hfill\vline\hfill
  \begin{minipage}[t]{0.49\linewidth}
    \[
      \small\begin{array}{l}
        \kw{WasteMakesHaste} :                                                     \\
        \quad\kwlet\ (I_1, I_1') = \ediv{\kwcake}{\emark{1}{0}{1 / 3}}\ \kwin\     \\
        \quad\kwlet\ (I_2, I_3) = \ediv{I_1'}{\emark{1}{0}{2 / 3 }}\ \kwin\        \\
        \quad\kwlet\ (A, B, C) = \kwsort_2(I_1, I_2, I_3)\ \kwin\                  \\
        \quad\kwif\ \eeval{2}{A}\ = \eeval{2}{B}\ \kwthen\                         \\
        \quad\quad \kwlet\ (A, B, C) = \kwsort_3(A, B, C)\ \kwin\                  \\
        \quad\quad \kwlet\ (B, C) = \kwsort_2(B, C)\ \kwin\                        \\
        \quad\quad \kwalloc(C, B, A)                                               \\
        \quad\kwelse\                                                              \\
        \quad\quad\kwlet\ (\mathit{Trim}, D) =                                     \\
        \quad\quad\quad\kwif\ (\eeval{2}{A} \geq 2 * \eeval{2}{B})\ \kwthen\       \\
        \quad\quad\quad\quad \ediv{A}{\emark{2}{0}{(1 / 2) * \eeval{2}{A}}}                        \\
        \quad\quad\quad\kwelse\                                                    \\
        \quad\quad\quad\quad \ediv{A}{\emark{2}{ \kwlt\ A}{\eeval{2}{B} }}         \\
        \quad\quad\kwin\                                                           \\
        \quad\quad\kwlet\ (A, B, C, D) = \kwsort_3(\mathit{Trim}, B, C, D)\ \kwin\ \\
        \quad\quad\kwif\ A = \mathit{Trim}\ \kwthen\                               \\
        \quad\quad\quad\kwlet\ (B, C, D) = \kwsort_2(B, C, D)\ \kwin\              \\
        \quad\quad\quad\kwlet\ (C, D) = \kwsort_1(C, D)\ \kwin\                    \\
        \quad\quad\quad\kwalloc(C, B, A)                                           \\
        \quad\quad\kwelse\                                                         \\
        \quad\quad\quad\kwlet\ (B, C, D) = \kwsort_1(B, C, D)\ \kwin\              \\
        \quad\quad\quad\kwif\ B = \mathit{Trim}\ \kwthen\                          \\
        \quad\quad\quad\quad \kwalloc(C, \mathit{Trim}, A)                         \\
        \quad\quad\quad\kwelse\                                                    \\
        \quad\quad\quad\quad \kwalloc(B, \mathit{Trim}, A)
      \end{array}\normalsize
    \]

    \vfill

    \includegraphics[width=\linewidth, height=4cm]{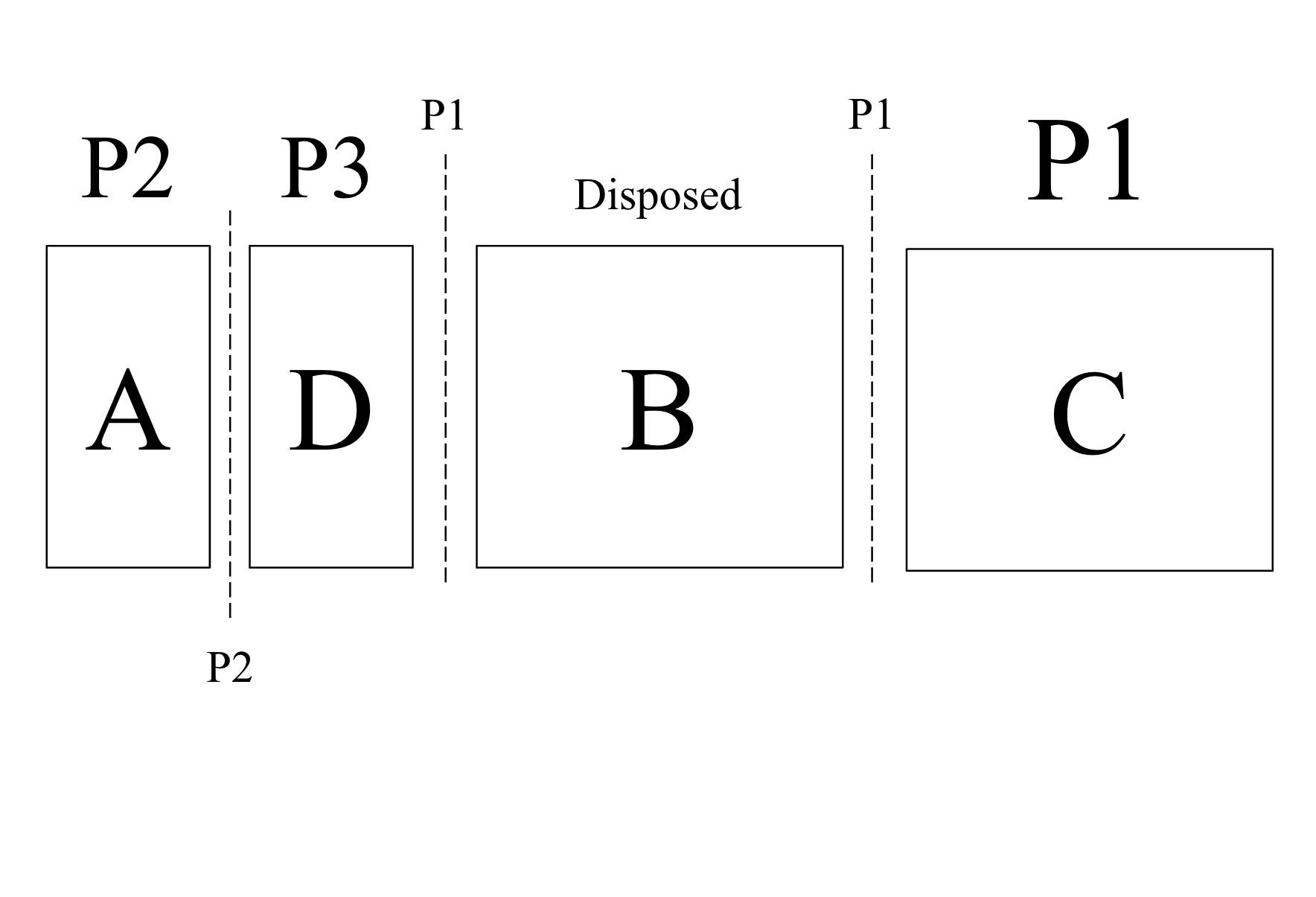}
    \begin{center}\text{Possible allocation from Waste-Makes-Haste}\end{center}

    \caption{Waste-Makes-Haste in \SYSTEM}
    \label{proto:wmh}
  \end{minipage}
\end{figure}

The Selfridge-Conway protocol achieves an envy-free allocation, but an agent's
allocation may be two disconnected pieces. This can be undesirable, for
instance if the agents are being allocated a stretch of road to build a store.
\citet{DBLP:journals/talg/Segal-HaleviHA16} propose a three-agent protocol
called Waste-Makes-Haste in the free-disposal model, where each agent receives
a \emph{contiguous} piece. The first phase of the protocol is the same as in
Selfridge-Conway, except the way P2 trims their favorite piece is different.
Then, each agent selects their favorite piece in a single round with a
prescribed order, rather than selecting two pieces in two rounds.
\Cref{proto:wmh} gives an idea of what the implementation looks like, along
with one possible allocation.

\section{Implementation and evaluation}
\label{sec:impl}

\subsection{System overview}
We have developed a prototype implementation of \SYSTEM in OCaml, using
Z3~\citep{DBLP:conf/tacas/MouraB08} and CVC5~\citep{cvc5} for constraint
solving. Our implementation is about 2000 lines of code. At a high level, our
tool converts fair-division protocols written in our domain-specific language
into logical constraints encoding properties like envy-freeness and
error-freeness, which are dispatched to Z3 and CVC5. We have also implemented an
evaluator for our language, so that users can run these verified protocols on
concrete valuations.

\paragraph{Front-end}
The surface language is quite similar to the core language described in
\Cref{sec:language}, with some syntactic sugar and an abbreviation facility for
writing common subroutines, like sorting pieces in descending order of
preference. Our implementation also includes a simple type-checker.

\paragraph{Evaluator}
Our evaluator implements the operational semantics in \Cref{rules}. The program
behavior depends on the agent valuations, and the evaluator has the following
type signature:
\[
  \mathtt{evaluate}:\ \mathtt{Protocol}\ \rightarrow\ \mathtt{Valuation}\ \mathtt{list}\ \rightarrow\ \mathtt{Mark}\ \mathtt{list}\ \rightarrow\ \mathtt{Alloc}
\]
Along with with the protocol to execute, the evaluator takes two arguments: a
list of valuation functions, and a list of marking functions. These two
parameters determine how $\kweval$ and $\kwmark$ queries behave, respectively.
While it would be simpler if we could provide just the valuation function, it is
not possible to derive the marking function from the valuation function without
further assumptions on the form of the valuation function. Our evaluator is
agnostic to how valuations and marks are implemented. In practice, the full
valuation and marking functions might not even be explicitly supplied; instead,
executing the protocol might involve querying actual agents how to value or mark
a given piece. Finally, the evaluator returns an allocation: a tuple of tuples
of intervals, describing which intervals each agent receives.

\paragraph{Valuation functions}
Our implementation treats valuations as functions from intervals to numbers.
This is not a limitation, because protocols make only a finite number of
cuts so each agent receives only finitely many disjoint intervals, and
valuations are assumed to be additive.
We list the solver axioms used for valuations in \append{app:impl}.

\paragraph{Constraint generation}
Our implementation converts the abstract syntax into logical formulas, which
are then dispatched to Z3 and CVC5. We model valuations as functions from
$\mathbb{R}^{2}$ to $\mathbb{R}$, and we encode our assumptions on valuations
as logical formulas that hold whenever the valuation is applied to two
arguments $(l, r)$ where $l \leq r$.

\paragraph{Verifying program properties}
As discussed at the end of \Cref{sec:constraints}, we can show that the output
of a protocol $e$ satisfies some property $\varphi(\ret)$ by showing that
\[
  \forall \ret. (\exists y_1,\ldots ,y_{\n{e}}.c(e,\ret))\Rightarrow \varphi(\ret)
\]
is valid, or equivalently, that
\begin{equation}
  \label{eq:notsat}
  \exists \ret. (\exists y_1,\ldots ,y_{\n{e}}.c(e,\ret))\wedge \neg \varphi(\ret)
\end{equation}
is not satisfiable. Once we generate the constraint of the protocol, we
pass~\eqref{eq:notsat} to Z3 and CVC5 to check for satisfiability. We apply this
technique for verifying envy-freeness to the formula representing envy-freeness
from the end of \Cref{sec:constraints}.



\subsection{Evaluation}

\begin{table}
  \caption{Verifying Envy-Freeness (timings averaged over $5$ runs)}
  \label{fig:bench}
  \begin{tabular}{@{}lrrrrr@{}}
    \toprule
    & \multicolumn{2}{c}{Size (Lines)} & \multicolumn{3}{c}{Timing (sec.)}\\
    \cmidrule{2-3}
    \cmidrule{4-6}
    \textbf{Protocol}        & \textbf{Program}     & \textbf{Constraint}   & \textbf{Compile} & \textbf{Z3} & \textbf{CVC5}  \\
    \midrule
    Cut-Choose               & 5    & 122   & 0.039      & 0.018  & 0.026\\
    Surplus                  & 5    & 132   & 0.042      & 0.020  & \text{failed}\\
    Waste-Makes-Haste        & 18   & 418   & 0.054      & 0.841  & 0.336\\
    Selfridge-Conway-Surplus & 20   & 401   & 0.051      & 0.819  & 0.573\\
    Selfridge-Conway-Full    & 22   & 786   & 0.103      & 19.376 & 53.46\\
    \bottomrule
  \end{tabular}
\end{table}

We implemented the protocols described in \Cref{sec:examples} and used our
implementation to verify envy-freeness. \Cref{fig:bench} shows basic metrics of
our benchmarks, and the time to verify each benchmark averaged over five trials
each. Experiments were performed on a machine with a 4 core Intel core i5 CPU
(4.5 GHz) with 16GB of \@RAM, running Linux. We have not explored how to
optimize our tool, so it is likely that performance can be further improved, but
our results show that our tool can verify fair-division protocols in a
reasonable amount of time. 

While it is difficult to measure the size of the generated constraints, our
rough metrics also show how the complexity of checking envy-freeness increases
rapidly for the more advanced protocols. The most complex protocol, the full
version of Selfridge-Conway, takes significantly more time to check than the
other protocols do under both Z3 and CVC5. We do not fully understand why this
protocol takes longer, but we believe that it is due to the number of paths
through this protocol.  Each path corresponds to a distinct disjunct in the
constraint. Computing the number of paths for our protocol, we find that the
Selfridge-Conway protocol has 1,800 paths, while Selfridge-Conway-Surplus and
Waste-Makes-Haste have the next highest path count at 216.




\section{Related work}
\label{sec:rw}

\paragraph{Alternative query models.}
Our work focuses on protocols in the Robertson-Webb query model. While this is
a quite common model in the fair-division literature, there are some protocols
that require other query models. For instance, some protocols ask the agents to
move multiple knives smoothly over the cake, until one agent decides to
cut~\citep{https://doi.org/10.48550/arxiv.1705.02946}. We believe that our
technical development can be extended with only minor modifications to handle
other query models, but it could be interesting to consider a more general
framework that can easily move between different models.

Another recent model is the generalized cut-choose model~\citep{AAAI1612294},
which allows agents to either make a new cut in the cake, or choose a single
piece in-between cuts to keep. This model is geared more towards cake division
viewed in a game-theoretic setting, where agents may deviate from the protocol.
While supporting new queries is fairly straightforward, reasoning about
correctness when agents may deviate is much more challenging---properties like
envy-freeness would not hold unconditionally, but could require assumptions
about the rationality of the agents. Incorporating such assumptions into
program verification is a fascinating direction for future work.

Researchers have also studied other models that limit the complexity of the
protocols. For instance, in the Simultaneous model, introduced by
\citet{bbkp2014}, the protocol asks each agent to report a compressed version
of their valuation, and is not allowed to make any further queries. These
models are useful for studying complexity properties of fair division
protocols. While our focus is on verifying envy-freeness, it would be
interesting to see if ideas from program verification, like resource analysis,
could be applied.

\paragraph{Verified mechanism design.}
There is a small, but growing area of formal verification aimed at programs from
the mechanism design literature. \citet{BGGHRS15} develop a type system for
relational properties, and show how to use their system to verify incentive
properties of auctions. \citet{HKM-verif16} later extended this method to verify
Bayesian incentive compatibility, a more sophisticated version of incentive
compatibility, for a randomized mechanism. Since envy-freeness does not appear
to be a relational property, it is unclear how to use existing methods to
analyze our target protocols.

While our work aims at fully automatic verification, researchers have
formalized various economic mechanisms and game-theoretic properties in
interactive theorem provers. For instance,
\citet{DBLP:conf/facs2/BaiTPG13,caminati2015sound} formalize incentive
properties of the VCG mechanism in Coq, and while developed a Coq library for
algorithmic game theory~\citep{DBLP:journals/jfrea/BagnallMS17} and mechanized
a proof that certain online learning algorithms converge to approximate
equilibria~\citep{DBLP:conf/esop/MertenBS18}.

\paragraph{Solver-aided programming.}
Finally, on the programming-language side, our work falls under the general
umbrella of solver-aided programming languages, which translate programs to
logical formulas, and then express target verification or synthesis goals as
logical constraints, which can be dispatched by automated SMT solvers. Some
prominent examples of tools in this area are Dafny~\citep{leino2010dafny},
F*~\citep{swamy2013verifying}, and Rosette~\citep{DBLP:conf/oopsla/TorlakB13}.
While it might be possible to implement our method using one of these systems,
it is not clear how to handle non-deterministic queries like $\kwmark$. Our protocols and properties are also
higher-order, since protocols depend on unknown valuation functions, and the
valuation functions are assumed to satisfy complex quantified axioms. It is not
clear how to encode our protocols and properties into existing solver-aided
tools. Additionally, since our programs are fairly simple, we do not need the
full generality of a solver-aided programming language. By developing our own,
more restricted language, we are able to prove soundness and completeness for
our constraint generation procedure and we are better able to tailor the
constraint generation for our target application.

\section{Conclusion and future directions}
\label{sec:conc}

In this work, we have developed a core language for fair-division protocols,
along with a constraint-generation procedure to encode envy-freeness and
error-freeness properties. Our prototype implementation \SYSTEM shows that it
is feasible to encode and check properties of these protocols
automatically. We see some natural directions for future work.

\paragraph{Verifying disjointness.}
Envy-freeness is the typical target property in fair-division, but there are
many other correctness properties that are not always obvious. For instance, a
protocol should always produce an allocation: a disjoint partition of the whole
cake, or of a subset of the cake if free disposal is allowed. It could be
interesting to develop a linear type system~\citep{DBLP:journals/tcs/Girard87}
to ensure that no part of the cake is allocated more than once; an affine type
system might be useful for allowing free disposal.

\paragraph{Developing custom solvers.}
So far, we have relied on general purpose SMT solvers (Z3 \& CVC5) to solve our
constraints. However, the constraints needed for our protocols are highly
specific to fair division, and it could be interesting to target constraint
generation to more specialized solvers, or even develop custom solvers for our
constraints. For instance, prior work on set constraint solvers (e.g.,
~\citep{DBLP:journals/scp/Aiken99}) might be useful in our setting.

\paragraph{Combinators for fair division.}
Finally, while we have made verification of protocols automatic, there is still
a major bottleneck: implementing the protocols in the first place. Converting
the existing presentation of advanced fair division protocols (say, for four or
more agents) into our language is extremely time consuming, often requiring
expert knowledge to resolve ambiguities and expand out the protocol from its
English description. It would be highly useful to extract patterns for
building fair division protocols; we could envision using these combinators to
structure the algorithms, making it easier to describe and implement protocols,
reducing code duplication, and improving constraint generation. Through our
initial experiments we have identified some possible combinators, but we
believe that much remains to be explored here.

\section*{Implementation access}
We have made available our prototype implementation online \citep{anonymous_2023_7806738}. It includes the implementation source, implementations of the examples from \Cref{sec:examples}, as well as scripts and step-by-step instructions that assist with reproducing Table \ref{fig:bench} and using our tool for other protocols.

\begin{acks}
  We thank the anonymous reviewers and the shepherd for their close reading and
  helpful suggestions. This work benefited from several rounds of feedback from
  the PL Discussion Group (PLDG) at Cornell. This work was partially supported
  by Cornell University and the NSF (Award \#1943130).
\end{acks}

\bibliographystyle{ACM-Reference-Format}
\bibliography{header,paper}


\newcommand{\SortNoop}[1]{}
\begin{thebibliography}{23}


\ifx \showCODEN    \undefined \def \showCODEN     #1{\unskip}     \fi
\ifx \showDOI      \undefined \def \showDOI       #1{#1}\fi
\ifx \showISBNx    \undefined \def \showISBNx     #1{\unskip}     \fi
\ifx \showISBNxiii \undefined \def \showISBNxiii  #1{\unskip}     \fi
\ifx \showISSN     \undefined \def \showISSN      #1{\unskip}     \fi
\ifx \showLCCN     \undefined \def \showLCCN      #1{\unskip}     \fi
\ifx \shownote     \undefined \def \shownote      #1{#1}          \fi
\ifx \showarticletitle \undefined \def \showarticletitle #1{#1}   \fi
\ifx \showURL      \undefined \def \showURL       {\relax}        \fi
\providecommand\bibfield[2]{#2}
\providecommand\bibinfo[2]{#2}
\providecommand\natexlab[1]{#1}
\providecommand\showeprint[2][]{arXiv:#2}

\bibitem[Aiken(1999)]%
        {DBLP:journals/scp/Aiken99}
\bibfield{author}{\bibinfo{person}{Alexander Aiken}.} \bibinfo{year}{1999}\natexlab{}.
\newblock \showarticletitle{Introduction to Set Constraint-Based Program Analysis}.
\newblock \bibinfo{journal}{\emph{Science of Computer Programming}} \bibinfo{volume}{35}, \bibinfo{number}{2} (\bibinfo{year}{1999}), \bibinfo{pages}{79--111}.
\newblock
\urldef\tempurl%
\url{https://doi.org/10.1016/S0167-6423(99)00007-6}
\showDOI{\tempurl}


\bibitem[Aziz and Mackenzie(2016a)]%
        {aziz2016discreteany}
\bibfield{author}{\bibinfo{person}{Haris Aziz} {and} \bibinfo{person}{Simon Mackenzie}.} \bibinfo{year}{2016}\natexlab{a}.
\newblock \showarticletitle{A discrete and bounded envy-free cake cutting protocol for any number of agents}. In \bibinfo{booktitle}{\emph{{IEEE} {S}ymposium on {F}oundations of {C}omputer {S}cience (FOCS), New Brunswick, New Jersey}}. \bibinfo{pages}{416--427}.
\newblock
\urldef\tempurl%
\url{https://doi.org/10.1109/FOCS.2016.52}
\showDOI{\tempurl}


\bibitem[Aziz and Mackenzie(2016b)]%
        {aziz2016discrete}
\bibfield{author}{\bibinfo{person}{Haris Aziz} {and} \bibinfo{person}{Simon Mackenzie}.} \bibinfo{year}{2016}\natexlab{b}.
\newblock \showarticletitle{A discrete and bounded envy-free cake cutting protocol for four agents}. In \bibinfo{booktitle}{\emph{{ACM} {SIGACT} {S}ymposium on {T}heory of {C}omputing (STOC), Cambridge, Massachusetts}}. \bibinfo{pages}{454--464}.
\newblock
\urldef\tempurl%
\url{https://doi.org/10.1145/2897518.2897522}
\showDOI{\tempurl}


\bibitem[Bagnall et~al\mbox{.}(2017)]%
        {DBLP:journals/jfrea/BagnallMS17}
\bibfield{author}{\bibinfo{person}{Alexander Bagnall}, \bibinfo{person}{Samuel Merten}, {and} \bibinfo{person}{Gordon Stewart}.} \bibinfo{year}{2017}\natexlab{}.
\newblock \showarticletitle{A Library for Algorithmic Game Theory in Ssreflect/Coq}.
\newblock \bibinfo{journal}{\emph{Journal of Formalized Reasoning}} \bibinfo{volume}{10}, \bibinfo{number}{1} (\bibinfo{year}{2017}), \bibinfo{pages}{67--95}.
\newblock
\urldef\tempurl%
\url{https://doi.org/10.6092/issn.1972-5787/7235}
\showDOI{\tempurl}


\bibitem[Bai et~al\mbox{.}(2013)]%
        {DBLP:conf/facs2/BaiTPG13}
\bibfield{author}{\bibinfo{person}{Wei Bai}, \bibinfo{person}{Emmanuel~M. Tadjouddine}, \bibinfo{person}{Terry~R. Payne}, {and} \bibinfo{person}{Sheng{-}Uei Guan}.} \bibinfo{year}{2013}\natexlab{}.
\newblock \showarticletitle{A Proof-Carrying Code Approach to Certificate Auction Mechanisms}. In \bibinfo{booktitle}{\emph{{F}ormal {A}aspects of {C}omponent {S}oftware (FACS), Nanchang, China}} \emph{(\bibinfo{series}{Lecture Notes in Computer Science}, Vol.~\bibinfo{volume}{8348})}, \bibfield{editor}{\bibinfo{person}{Jos{\'{e}}~Luiz Fiadeiro}, \bibinfo{person}{Zhiming Liu}, {and} \bibinfo{person}{Jinyun Xue}} (Eds.). \bibinfo{publisher}{Springer}, \bibinfo{pages}{23--40}.
\newblock
\urldef\tempurl%
\url{https://doi.org/10.1007/978-3-319-07602-7\_4}
\showDOI{\tempurl}


\bibitem[Balkanski et~al\mbox{.}(2014)]%
        {bbkp2014}
\bibfield{author}{\bibinfo{person}{Eric Balkanski}, \bibinfo{person}{Simina Brânzei}, \bibinfo{person}{David Kurokawa}, {and} \bibinfo{person}{Ariel Procaccia}.} \bibinfo{year}{2014}\natexlab{}.
\newblock \showarticletitle{Simultaneous Cake Cutting}.
\newblock \bibinfo{journal}{\emph{{AAAI} Conference on Artificial Intelligence, Québec, Canada}} \bibinfo{volume}{28}, \bibinfo{number}{1} (\bibinfo{date}{Jun.} \bibinfo{year}{2014}).
\newblock
\urldef\tempurl%
\url{https://doi.org/10.1609/aaai.v28i1.8802}
\showDOI{\tempurl}


\bibitem[Barbosa et~al\mbox{.}(2022)]%
        {cvc5}
\bibfield{author}{\bibinfo{person}{Haniel Barbosa}, \bibinfo{person}{Clark~W. Barrett}, \bibinfo{person}{Martin Brain}, \bibinfo{person}{Gereon Kremer}, \bibinfo{person}{Hanna Lachnitt}, \bibinfo{person}{Makai Mann}, \bibinfo{person}{Abdalrhman Mohamed}, \bibinfo{person}{Mudathir Mohamed}, \bibinfo{person}{Aina Niemetz}, \bibinfo{person}{Andres N{\"{o}}tzli}, \bibinfo{person}{Alex Ozdemir}, \bibinfo{person}{Mathias Preiner}, \bibinfo{person}{Andrew Reynolds}, \bibinfo{person}{Ying Sheng}, \bibinfo{person}{Cesare Tinelli}, {and} \bibinfo{person}{Yoni Zohar}.} \bibinfo{year}{2022}\natexlab{}.
\newblock \showarticletitle{cvc5: {A} Versatile and Industrial-Strength {SMT} Solver}. In \bibinfo{booktitle}{\emph{International Conference on Tools and Algorithms for the Construction and Analysis of Systems (TACAS), Munich, Germany}} \emph{(\bibinfo{series}{Lecture Notes in Computer Science}, Vol.~\bibinfo{volume}{13243})}, \bibfield{editor}{\bibinfo{person}{Dana Fisman} {and} \bibinfo{person}{Grigore Rosu}} (Eds.). \bibinfo{publisher}{Springer}, \bibinfo{pages}{415--442}.
\newblock
\urldef\tempurl%
\url{https://doi.org/10.1007/978-3-030-99524-9\_24}
\showDOI{\tempurl}


\bibitem[Barthe et~al\mbox{.}(2015)]%
        {BGGHRS15}
\bibfield{author}{\bibinfo{person}{Gilles Barthe}, \bibinfo{person}{Marco Gaboardi}, \bibinfo{person}{Emilio~Jes{\'u}s Gallego~Arias}, \bibinfo{person}{Justin Hsu}, \bibinfo{person}{Aaron Roth}, {and} \bibinfo{person}{Pierre-Yves Strub}.} \bibinfo{year}{2015}\natexlab{}.
\newblock \showarticletitle{Higher-Order Approximate Relational Refinement Types for Mechanism Design and Differential Privacy}. In \bibinfo{booktitle}{\emph{{ACM} {SIGPLAN--SIGACT} {S}ymposium on {P}rinciples of {P}rogramming {L}anguages ({POPL}), Mumbai, India}}. \bibinfo{pages}{55--68}.
\newblock
\urldef\tempurl%
\url{https://doi.org/10.1145/2676726.2677000}
\showDOI{\tempurl}
\showeprint[arxiv]{1407.6845}~[cs.PL]


\bibitem[Barthe et~al\mbox{.}(2016)]%
        {HKM-verif16}
\bibfield{author}{\bibinfo{person}{Gilles Barthe}, \bibinfo{person}{Marco Gaboardi}, \bibinfo{person}{Emilio~Jes{\'u}s Gallego~Arias}, \bibinfo{person}{Justin Hsu}, \bibinfo{person}{Aaron Roth}, {and} \bibinfo{person}{Pierre-Yves Strub}.} \bibinfo{year}{2016}\natexlab{}.
\newblock \showarticletitle{Computer-Aided Verification in Mechanism Design}. In \bibinfo{booktitle}{\emph{Conference on Web and Internet Economics (WINE), Montr{\'e}al, Qu{\'e}bec}} \emph{(\bibinfo{series}{Lecture Notes in Computer Science}, Vol.~\bibinfo{volume}{10123})}. \bibinfo{publisher}{Springer-Verlag}, \bibinfo{pages}{273--293}.
\newblock
\urldef\tempurl%
\url{https://doi.org/10.1007/978-3-662-54110-4_20}
\showDOI{\tempurl}
\showeprint[arxiv]{1502.04052}~[cs.GT]


\bibitem[Bertram et~al\mbox{.}(2023)]%
        {anonymous_2023_7806738}
\bibfield{author}{\bibinfo{person}{Noah Bertram}, \bibinfo{person}{Alex Levinson}, {and} \bibinfo{person}{Justin Hsu}.} \bibinfo{year}{2023}\natexlab{}.
\newblock \bibinfo{booktitle}{\emph{Cutting the Cake: A Language for Fair Division}}.
\newblock
\urldef\tempurl%
\url{https://doi.org/10.5281/zenodo.7806738}
\showDOI{\tempurl}


\bibitem[Brams et~al\mbox{.}(2006)]%
        {brams2006better}
\bibfield{author}{\bibinfo{person}{Steven~J Brams}, \bibinfo{person}{Michael~A Jones}, {and} \bibinfo{person}{Christian Klamler}.} \bibinfo{year}{2006}\natexlab{}.
\newblock \showarticletitle{Better Ways to Cut a Cake}.
\newblock \bibinfo{journal}{\emph{Notices of the {AMS}}} \bibinfo{volume}{53}, \bibinfo{number}{11} (\bibinfo{year}{2006}), \bibinfo{pages}{1314--1321}.
\newblock


\bibitem[Brânzei et~al\mbox{.}(2016)]%
        {AAAI1612294}
\bibfield{author}{\bibinfo{person}{Simina Brânzei}, \bibinfo{person}{Ioannis Caragiannis}, \bibinfo{person}{David Kurokawa}, {and} \bibinfo{person}{Ariel Procaccia}.} \bibinfo{year}{2016}\natexlab{}.
\newblock \showarticletitle{An Algorithmic Framework for Strategic Fair Division}.
\newblock \bibinfo{journal}{\emph{{AAAI} Conference on Artificial Intelligence, Phoenix, Arizona}} (\bibinfo{year}{2016}).
\newblock
\urldef\tempurl%
\url{https://doi.org/10.1609/aaai.v30i1.10042}
\showDOI{\tempurl}


\bibitem[Brânzei and Nisan(2017)]%
        {https://doi.org/10.48550/arxiv.1705.02946}
\bibfield{author}{\bibinfo{person}{Simina Brânzei} {and} \bibinfo{person}{Noam Nisan}.} \bibinfo{year}{2017}\natexlab{}.
\newblock \bibinfo{title}{The Query Complexity of Cake Cutting}.
\newblock
\newblock
\urldef\tempurl%
\url{https://doi.org/10.48550/ARXIV.1705.02946}
\showDOI{\tempurl}


\bibitem[Caminati et~al\mbox{.}(2015)]%
        {caminati2015sound}
\bibfield{author}{\bibinfo{person}{Marco~B Caminati}, \bibinfo{person}{Manfred Kerber}, \bibinfo{person}{Christoph Lange}, {and} \bibinfo{person}{Colin Rowat}.} \bibinfo{year}{2015}\natexlab{}.
\newblock \showarticletitle{Sound auction specification and implementation}. In \bibinfo{booktitle}{\emph{{ACM} {SIGecom} {C}onference on {E}conomics and {C}omputation (EC), Portland, Oregon}}. \bibinfo{pages}{547--564}.
\newblock
\urldef\tempurl%
\url{https://doi.org/10.1145/2764468.2764511}
\showDOI{\tempurl}


\bibitem[de~Moura and Bj{\o}rner(2008)]%
        {DBLP:conf/tacas/MouraB08}
\bibfield{author}{\bibinfo{person}{Leonardo~Mendon{\c{c}}a de Moura} {and} \bibinfo{person}{Nikolaj~S. Bj{\o}rner}.} \bibinfo{year}{2008}\natexlab{}.
\newblock \showarticletitle{{Z3:} An Efficient {SMT} Solver}. In \bibinfo{booktitle}{\emph{International Conference on Tools and Algorithms for the Construction and Analysis of Systems (TACAS), Budapest, Hungary}} \emph{(\bibinfo{series}{Lecture Notes in Computer Science}, Vol.~\bibinfo{volume}{4963})}, \bibfield{editor}{\bibinfo{person}{C.~R. Ramakrishnan} {and} \bibinfo{person}{Jakob Rehof}} (Eds.). \bibinfo{publisher}{Springer}, \bibinfo{pages}{337--340}.
\newblock
\urldef\tempurl%
\url{https://doi.org/10.1007/978-3-540-78800-3\_24}
\showDOI{\tempurl}


\bibitem[Girard(1987)]%
        {DBLP:journals/tcs/Girard87}
\bibfield{author}{\bibinfo{person}{Jean{-}Yves Girard}.} \bibinfo{year}{1987}\natexlab{}.
\newblock \showarticletitle{Linear Logic}.
\newblock \bibinfo{journal}{\emph{Theoretical Computer Science}}  \bibinfo{volume}{50} (\bibinfo{year}{1987}), \bibinfo{pages}{1--102}.
\newblock
\urldef\tempurl%
\url{https://doi.org/10.1016/0304-3975(87)90045-4}
\showDOI{\tempurl}


\bibitem[Leino(2010)]%
        {leino2010dafny}
\bibfield{author}{\bibinfo{person}{K~Rustan~M Leino}.} \bibinfo{year}{2010}\natexlab{}.
\newblock \showarticletitle{Dafny: An automatic program verifier for functional correctness}. In \bibinfo{booktitle}{\emph{International Conference on Logic for Programming, Artificial Intelligence and Reasoning (LPAR), Senegal, Dakar}}. Springer, \bibinfo{pages}{348--370}.
\newblock


\bibitem[Merten et~al\mbox{.}(2018)]%
        {DBLP:conf/esop/MertenBS18}
\bibfield{author}{\bibinfo{person}{Samuel Merten}, \bibinfo{person}{Alexander Bagnall}, {and} \bibinfo{person}{Gordon Stewart}.} \bibinfo{year}{2018}\natexlab{}.
\newblock \showarticletitle{Verified Learning Without Regret - From Algorithmic Game Theory to Distributed Systems with Mechanized Complexity Guarantees}. In \bibinfo{booktitle}{\emph{European Symposium on Programming (ESOP), Thessaloniki, Greece}} \emph{(\bibinfo{series}{Lecture Notes in Computer Science}, Vol.~\bibinfo{volume}{10801})}, \bibfield{editor}{\bibinfo{person}{Amal Ahmed}} (Ed.). \bibinfo{publisher}{Springer-Verlag}, \bibinfo{pages}{561--588}.
\newblock
\urldef\tempurl%
\url{https://doi.org/10.1007/978-3-319-89884-1\_20}
\showDOI{\tempurl}


\bibitem[Robertson and Webb(1998)]%
        {robertson1998cake}
\bibfield{author}{\bibinfo{person}{Jack Robertson} {and} \bibinfo{person}{William Webb}.} \bibinfo{year}{1998}\natexlab{}.
\newblock \bibinfo{booktitle}{\emph{Cake-cutting algorithms: Be fair if you can}}.
\newblock \bibinfo{publisher}{AK Peters/CRC Press}.
\newblock


\bibitem[Segal{-}Halevi et~al\mbox{.}(2016)]%
        {DBLP:journals/talg/Segal-HaleviHA16}
\bibfield{author}{\bibinfo{person}{Erel Segal{-}Halevi}, \bibinfo{person}{Avinatan Hassidim}, {and} \bibinfo{person}{Yonatan Aumann}.} \bibinfo{year}{2016}\natexlab{}.
\newblock \showarticletitle{Waste Makes Haste: Bounded Time Algorithms for Envy-Free Cake Cutting with Free Disposal}.
\newblock \bibinfo{journal}{\emph{ACM Transactions on Algorithms}} \bibinfo{volume}{13}, \bibinfo{number}{1} (\bibinfo{year}{2016}), \bibinfo{pages}{12:1--12:32}.
\newblock
\urldef\tempurl%
\url{https://doi.org/10.1145/2988232}
\showDOI{\tempurl}


\bibitem[Swamy et~al\mbox{.}(2013)]%
        {swamy2013verifying}
\bibfield{author}{\bibinfo{person}{Nikhil Swamy}, \bibinfo{person}{Juan Chen}, {and} \bibinfo{person}{Ben Livshits}.} \bibinfo{year}{2013}\natexlab{}.
\newblock \showarticletitle{Verifying Higher-order Programs with the Dijkstra Monad}. In \bibinfo{booktitle}{\emph{{ACM SIGPLAN Conference on Programming Language Design and Implementation (PLDI)}, Seattle, Washington}}. \bibinfo{publisher}{ACM}.
\newblock
\urldef\tempurl%
\url{https://doi.org/10.1145/2499370.2491978}
\showDOI{\tempurl}


\bibitem[Torlak and Bod{\'{\i}}k(2013)]%
        {DBLP:conf/oopsla/TorlakB13}
\bibfield{author}{\bibinfo{person}{Emina Torlak} {and} \bibinfo{person}{Rastislav Bod{\'{\i}}k}.} \bibinfo{year}{2013}\natexlab{}.
\newblock \showarticletitle{Growing solver-aided languages with rosette}. In \bibinfo{booktitle}{\emph{{ACM} International Symposium on New ideas, new paradigms, and reflections on programming \& software (Onward!), Indianapolis, Indiana}}, \bibfield{editor}{\bibinfo{person}{Antony~L. Hosking}, \bibinfo{person}{Patrick~Th. Eugster}, {and} \bibinfo{person}{Robert Hirschfeld}} (Eds.). \bibinfo{publisher}{{ACM}}, \bibinfo{pages}{135--152}.
\newblock
\urldef\tempurl%
\url{https://doi.org/10.1145/2509578.2509586}
\showDOI{\tempurl}


\bibitem[Woeginger and Sgall(2007)]%
        {WOEGINGER2007213}
\bibfield{author}{\bibinfo{person}{Gerhard~J. Woeginger} {and} \bibinfo{person}{Jiří Sgall}.} \bibinfo{year}{2007}\natexlab{}.
\newblock \showarticletitle{On the complexity of cake cutting}.
\newblock \bibinfo{journal}{\emph{Discrete Optimization}} \bibinfo{volume}{4}, \bibinfo{number}{2} (\bibinfo{year}{2007}), \bibinfo{pages}{213--220}.
\newblock
\showISSN{1572-5286}
\urldef\tempurl%
\url{https://doi.org/10.1016/j.disopt.2006.07.003}
\showDOI{\tempurl}


\end{thebibliography}

\appendix
\iflong
  \section{Omitted Proofs}
  \label{app:proofs}
  Here we detail proofs from \Cref{sec:constraints}.
  To work up to the main proofs, we will need a few technical lemmas. First, the
  constraint translation only creates logical variables between $y_{k+1}$ and
  $y_{k + \n{e}}$.
  \begin{lemma}
    Let $e$ be any expression. For any $k \in \mathbb{N}$, we have $\emap_{Q}(k,
      e) = n(e) + k$ and
    \begin{equation*}
      \fv(c(k, e, \ret))\cap \mathcal{Y} = \fv(\emap_{s}(k,e)) \cap \mathcal{Y}
      = \fv(\emap_{\rho}(k,e)) \cap \mathcal{Y} = \{y_{k + 1},\ldots ,y_{k + n(e)}\} .
    \end{equation*}
    \label{lem:yinc}
  \end{lemma}

  Thus, $\n{e} = |\fv(c(k,e,\ret))\cap \mathcal{Y}|$ for any $k$. Second, the
  free variables of $e$ are precisely the logical variables in $\mathcal{X}$
  appearing in the constraint of $e$.

  \begin{lemma}
    \label{lem:xinc}
    $\fv(e) = \fv(c(k, e, \ret))\cap \mathcal{X}$ for any $k \in \mathbb{N}$.
  \end{lemma}

  Both lemmas follow by induction on $e$. Next, if the domain of a substitution
  $S$ covers all free variables in $e$, then $S$ substitutes away all program
  variables in the constraint of $e$. Note that this is not immediate---a
  substitution replaces variables by logical terms, which might contain other
  variables---but the structure of substitutions ensures that all program
  variables are replaced.

  \begin{lemma}
    \label{sublemma}
    Let $S$ be a substitution such that $\fv(e)\subseteq \dom(S)$. Then
    \[
      \fv(c(k, e,\ret)S)\cap\mathcal{X} = \emptyset.
    \]
  \end{lemma}
  \begin{proof}
    Let $S = \{t_{n}/x_{n}\}\cdots \{t_1/x_1\}$.  We show the more general fact if
    $\varphi$ is a formula such that $\fv(\varphi)\cap \mathcal{X} \subseteq
      \dom(S)$, then
    \begin{equation}
      \fv(\varphi\{t_{n}/ \vemb{x_{n}}\}\cdots \{t_{n-i + 1}/ \vemb{x_{n -i + 1}}\})\cap \mathcal{X}\subseteq \{x_1,\ldots x_{n-i}\}
    \end{equation}
    for $i = 0,\ldots ,n$
    by induction on $i$.
    The base case follows by definition:
    \[
      \fv(\varphi)\cap \mathcal{X} \subseteq \dom(S)= \{x_1,\ldots ,x_{n}\}.
    \]
    Suppose the hypothesis holds for $i-1$, then we have that
    \begin{align*}
       & \fv(\varphi\{t_{n}/ \vemb{x_{n}}\}\cdots \{t_{n-(i-1) + 1}/ \vemb{x_{n -(i-1) + 1}}\}\{t_{n-i + 1}/ \vemb{x_{n -i +1}}\})\cap \mathcal{X}                                      \\
       & = \fv(\varphi\{t_{n}/ \vemb{x_{n}}\}\cdots \{t_{n-(i-1) + 1}/ \vemb{x_{n -(i-1) + 1}}\})\setminus\{x_{n-i + 1}\}\cup \fv(\{t_{n- i + 1}\})\cap \mathcal{X}                     \\
       & = (\fv(\varphi\{t_{n}/ \vemb{x_{n}}\}\cdots \{t_{n-(i-1) + 1}/ \vemb{x_{n -(i-1) + 1}}\})\cap \mathcal{X})\setminus\{x_{n-i + 1}\}\cup (\fv(\{t_{n- i + 1}\})\cap \mathcal{X}) \\
       & \subseteq \{x_1 ,\ldots ,x_{n - i +1}\}\setminus \{x_{n-i + 1}\} \cup
      \{x_1,\ldots ,x_{n-i}\}                                                                                                                                                           \\
       & = \{x_1,\ldots x_{n-i}\}
    \end{align*}
    as desired.
    Thus taking $i$ to be $n$, we conclude that
    \[
      \fv(\varphi\{t_{n}/ \vemb{x_{n}}\}\cdots \{t_{1}/ \vemb{x_{1}}\})\cap \mathcal{X}\subseteq \emptyset.
    \]
    Applying \Cref{lem:xinc}, our claim is shown.
  \end{proof}

  \soundness*
  \begin{proof}
    We proceed by induction on the derivation of $\left< e, \sigma
      \right>\Downarrow_{\vset} v$. Suppose that $\sigma$ and $S$ are both on
    $x_1,\ldots ,x_{n}$. We proceed by case analysis on the evaluation rule for
    $\left< e,\sigma \right> \Downarrow_{\vset} v$.
    \begin{description}
      \item[\rname{Val}]
        We have $c(k, e, \emb{v}) = c(k , v,\emb{v}) = (\emb{v} = \emb{v})$ for any
        $k$, so $\interp \vDash c(k, e,\emb{v})$.
      \item[\rname{Var}]
        We have $v = \sigma(x)$. Because $\interp$ marries $\sigma$ and $S$,
        $\interp(xS) = \sigma(x)$ and so
        \[
          \interp \vDash (\emb{\sigma(x)} = \vemb{x})S.
        \]
        Noting that $c(k, x,\emb{\sigma(x)}) = \mathsf{true}\wedge(\emb{\sigma(x)} =
          \vemb{x})$ finishes this case, setting $\interp' = \interp$.
      \item[\rname{Let}]
        We have $e = (\kwlet\ x = e_1\ \kwin\ e_2)$ such that
        $\left<e_1, \sigma \right>
          \Downarrow_{\vset} v_{1}$ and $\left< e_2,
          \sigma'\right>\Downarrow_{\vset} v$, where we set $\sigma' = \sigma
          [x\mapsto v_1]$. Because $\fv(e_1) \subseteq \fv(e)\subseteq \dom(\sigma)$
        and $\interp$ marries $\sigma$ and $S$, we can apply the inductive
        hypothesis to obtain an interpretation $\interp_1 \in \M_{\vset}$ such
        that
        \begin{equation*}
          \interp_1 \vDash c(k, e_1,\emb{v_1})S.
        \end{equation*}
        Unpacking, we have
        \[
          \interp_1 \vDash (\emap_{\rho}(k, e_1) = \emb{v_1})S,
        \]
        which can be expressed as
        \[
          \interp_1 (\vemb{x}S')= \sigma'(x),
        \]
        where we set $S' = \{\emap_{\rho}(k, e_1)/ \vemb{x}\}S$. This shows that
        $\interp_1 $ also marries $\sigma'$ and $S'$. Notice in addition that
        $\fv(e_2)\subseteq \dom(\sigma')$. Thus by induction, for $k' = k + \n{e_1} =
          m(S')$ (from \Cref{lem:yinc} and because $k\geq m(S)$), there is some proper
        interpretation $\interp'$ from $\M_{\vset}$ such that
        \begin{equation}
          \interp' \vDash  c(k', e_2, \emb{v})S',
          \label{e_2c}
        \end{equation}
        and $\interp'$ agrees with $\interp_1$ up to $k'$. By \Cref{lem:yinc},
        $\fv(c(k, e_1, \emb{v_1})S) \subseteq \{y_1 ,\ldots ,y_{k'}\}$, so we
        additionally obtain
        \begin{equation}
          \interp' \vDash c(k, e_1,\emb{v_1})S.
          \label{e_1c}
        \end{equation}
        Combining \eqref{e_1c} with \eqref{e_2c}, we have
        \[
          \interp' \vDash \emap_{s}(k, e_1)S\wedge \emap_s(k', e_2)S'\wedge(\emb{v}
          = \emap_{\rho}(k', e_2))S' .
        \]
        Recalling how we defined $k'$ and $S'$, we notice that the above is exactly
        \[
          \interp' \vDash c(k, e,\widehat{v})S.
        \]
        Lastly, because $\interp'$ agrees with $\interp_1$ up to $k'$ and $\interp_1$
        agrees with $\interp$ up to $k$, we conclude that $\interp'$ agrees with
        $\interp$ up to $k$, so we are done with this case.
      \item[\rname{Ops}]
        We have $v = \sem{o}(v_1,\ldots ,v_{n})$, $e = o(e_1 ,\ldots ,e_{n})$, and
        $\left< e_{i},\sigma \right>\Downarrow_{\vset} v_{i}$ for $i = 1 ,\ldots
          ,n$ for some $v_1,\ldots ,v_{n} \in \val$, $o \in \mathcal{O}$, and
        $e_1,\ldots ,e_{n}$. By induction, we have $\interp_{i} \in \M_{\vset}$
        such that
        \begin{equation}
          \interp_{i} \vDash  c( k + \n{e_1,\ldots ,e_{i -1}}, e_i, \emb{v_{i}})S
        \end{equation}
        and $\interp_{i}$ agrees with $\interp$ up to $k$ for all $i = 1,\ldots
          ,n$.  By \Cref{lem:yinc}, the variables from $\mathcal{Y}$ in all the
        constraints are pair-wise disjoint. Thus, we can obtain an interpretation
        $\interp'$ from $\M_{\vset}$ that agrees with $\interp$ up to $k$ for
        which
        \[
          \interp' \vDash c(k + \n{e_1,\ldots ,e_{i-1}} , e_{i}, \emb{v_{i}})S
        \]
        for $i = 1,\ldots ,n$. Recalling the definition of the constraint of $e$, we
        have $\interp'\vDash \emap_{s}(k + \n{e_1,\ldots ,e_{i -1}}, e_{i})S$ and
        $\interp'(\emap_{\rho}(k + \n{e_1,\ldots ,e_{i-1}}, e_{i})S) = v_{i}$ for $i =
          1,\ldots ,n$. Because $\interp'(o) = \sem{o}$ as $\interp'$ is proper, we also
        have
        \[
          \interp'(o(\emap_{\rho}(k, e_1) ,\ldots ,\emap_{\rho}(k + \n{e_1,\ldots ,e_{n-1}}, e_{n}))S)
          = \sem{o}(v_1,\ldots ,v_{n}) = v,
        \]
        so we can conclude that
        \[
          \interp' \vDash  c( k, e, \emb{v})S.
        \]
        The cases of \rname{Pair}, \rname{Proj}, \rname{Cake}, \rname{Lt}, \rname{Rt},
        and \rname{Eval} are all handled near identically to \rname{Ops} so we omit
        them.
      \item[\rname{IfTrue}]
        We have $e = (\kwif\ e_1\ \kwthen\ e_2\ \kwelse\ e_3)$,
        and $\left< e_1, \sigma \right> \Downarrow_{\vset} \mathsf{true}$,
        and $v = v_2$. By induction, there exists $\interp_1 \in \M_{\vset}$ such that
        \begin{equation*}
          \interp_1 \vDash c(k, e_1,\emb{\mathsf{true}})S
        \end{equation*}
        and $\interp_2 \in \M_{\vset}$ such that
        \begin{equation*}
          \interp_2 \vDash c(k + \n{e_1}, e_2,\emb{v_2})S
        \end{equation*}
        and both agree with $\interp$ up to $k$.
        We can construct an interpretation $\interp' \in
          \M_{\vset}$ by setting $\interp'(y_{j}) = \interp_1(y_{j})$ for $j =
          1,\ldots , k + \n{e_1}$, and $\interp'(y_{j}) = \interp_2(y_{j})$ for all
        other $j$. Using \Cref{lem:yinc}, we have
        \begin{equation}
          \interp' \vDash c(k, e_1, \emb{\mathsf{true}})S \wedge c(k,e_2,\emb{v_2} )S.
          \label{eq:tautwoprime}
        \end{equation}
        This means in particular that $\interp'(\emap_{\rho}(k, e_1)S) =
          \mathsf{true}$ and also that $\interp'(\emap_{\rho}(k + \n{e_1}, e_2)S) =
          v_2$. By the interpretation of $\ite{}{}{}$, we have
        \begin{equation}
          \interp' (\ite{\emap_{\rho}(k, e_1)}{\emap_{\rho}(k + \n{e_1}, e_2)}{\emap_{\rho}(k + \n{e_1, e_2}, e_3)}S) = \emb{v_2}.
          \label{eq:tautwoite}
        \end{equation}
        Using \eqref{eq:tautwoprime} again, we also have
        \begin{equation*}
          \interp' \vDash \emap_s(k, e_1)S
          \wedge \emap_s(k + \n{e_1}, e_2)S
          \wedge (\emap_{\rho}(k, e_1)S
          = \mathsf{true}),
        \end{equation*}
        which implies
        \begin{align*}
          \interp'  & \vDash (\emap_s(k, e_1)S\wedge \emap_s(k + \n{e_1}, e_2)S\wedge (\emap_{\rho}(k, e_1)S = \mathsf{true})) \\
          \nonumber & \qquad\vee (\emap_s(k, e_1)S\wedge \emap_s(k', e_3)S\wedge (\emap_{\rho}(k, e_1)S = \mathsf{false}))
        \end{align*}
        But now we can factor out $\emap_s(k, e_1)$ and $S$ to arrive at
        \begin{align*}
          \interp' & \vDash (\emap_s(k, e_1)                                                                     \\
                   & \qquad\wedge (\emap_s(k + \n{e_1}, e_2)\wedge (\emap_{\rho}(k, e_1) = \mathsf{true}))       \\
                   & \qquad\vee (\emap_s(k + \n{e_1,e_2}, e_3)\wedge (\emap_{\rho}(k, e_1) = \mathsf{false})))S.
          \label{eq:itec}
        \end{align*}
        Combining this with \eqref{eq:tautwoite}, we have
        \[
          \interp' \vDash c(k, e,\emb{v})S
        \]
        as desired.
      \item[\rname{IfFalse}]
        Similar to \rname{IfTrue}.
      \item[\rname{Mark}]
        We have $e = \emark{a}{e_1}{e_2}$ and $\left<e_1, \sigma \right>
          \Downarrow_{\vset} v_1$ and $\left< e_2, \sigma \right> \Downarrow_{\vset}
          v_2$, and $\left< e,\sigma \right>\Downarrow_{\vset} r$ for some $r
          \in \mathbb{R}$. Additionally, we have $V_{a}(\left[ v_1, 1
            \right])\geq v_{2}$ and $V_{a}(\left[v_1, r \right]) = v_2$. Using the
        inductive hypothesis, there exists $\interp_1$ and $\interp_2$ from
        $\M_{\vset}$ such that $\interp_1 \vDash c(k, e_1,\emb{v_1})S$ and
        $\interp_2 \vDash c(k + \n{e_1}, e_2,\emb{v_2})S$, for which both
        $\interp_1$ and $\interp_2$ agree with $\interp$ up to $k$. From here, we
        can see from \Cref{lem:yinc} that $\fv(c(k + \n{e_1}, e_2,\emb{v_2})) \cap
          \fv(c(k, e_1,\emb{v_1})) \cap \mathcal{Y} = \emptyset$. Therefore, using
        $\interp_1$ and $\interp_2$, we can construct $\interp'$ from
        $\M_{\vset}$ that agrees with $\interp$ up to $k$, and in particular
        $\interp'(y_{k'} ) = r$, where $k' = k + \n{e_1} + 1$, and
        \[
          \interp' \vDash
          \emap_s(k, e_1')S\wedge \emap_s(k + \n{e_1'}, e_2')S \wedge \v_{a}(\left[ v_1, 1 \right]\geq \emb{v_{2}}) \wedge(\v_{a}(\left[v_1, y_{k'} \right]) = \emb{v_2}) \wedge (\emb{v} = y_{k'}).
        \]
        Pulling out $S$, we have $\interp' \vDash c(k, e,\emb{v})S$ as desired.
      \item[\rname{Div}]
        We have $v = ([r_1,r_2], [r_2,r_1'])$, $r_1\leq r_2\leq r_1'$, $e = \ediv{e_1}{e_2}$, and
        both $\left< e_{1},\sigma \right>\Downarrow_{\vset} [r_1,r_1']$ and $\left< e_2,\sigma \right>\Downarrow_{\vset} r_2$.
        By induction, we have $\interp_{1}, \interp_2 \in \M_{\vset}$ such that
        \begin{equation}
          \interp_{1} \vDash  c( k, e_1, \emb{[r_1,r_1']})S\quad\text{and}\quad \interp_{2} \vDash  c( k + \n{e_1}, e_2, \emb{r_2})S
        \end{equation}
        and both $\interp_{1}$ and $\interp_2$ agree with $\interp$ on $\{y_1,\ldots ,y_{k}\}$. By \Cref{lem:yinc},
        \[
          \fv(c( k, e_1, \emb{[r_1,r_1']}))\cap \fv(c( k + \n{e_1}, e_2, \emb{r_2})) \cap \mathcal{Y} = \emptyset.
        \]
        Thus, we can obtain an interpretation $\interp'$ from $\M_{\vset}$ that agrees
        with $\interp$ on $\{y_1,\ldots ,y_{k}\}$ for which
        \[
          \interp' \vDash  c( k, e_1, \emb{[r_1,r_1']})S\wedge  c( k + \n{e_1}, e_2, \emb{r_2})S
        \]
        Recalling the definition of the constraint of $e$, we have $\interp'\vDash
          \emap_{s}(k,e_{1})S\wedge \emap_{s}(k + \n{e_1}, e_{2})S$ and
        $\interp'(\emap_{\rho}(k, e_{1})S) = [r_1,r_1']$ and $\interp'(\emap_{\rho}(k +
          \n{e_1}, e_{2})S)= r_2$. Because $\interp'(\ell)$ and $\interp'(r)$ gives the
        left and right endpoints of an interval respectively, we also have that
        $\interp'(\ell(\emap_{\rho}(k, e_{1})S)) = r_1$ and $\interp'(r(\emap_{\rho}(k,
          e_{1})S)) = r_1'$. Thus, because $r_1' \geq r_2 \geq r_1$ and using that $\mu'$
        is proper again, it is easy to see that
        \[
          \interp' \vDash  c( k, e, \emb{v})S.
        \]
        The cases of \rname{Pair}, \rname{Proj}, \rname{Cake}, \rname{Lt}, \rname{Rt},
        and \rname{Eval} are all handled near identically to \rname{Ops} so we omit
        them. \qedhere
    \end{description}
  \end{proof}
  \completeness*
  \begin{proof}
    We proceed by induction on the syntax of $e$. Suppose that $\interp \vDash c(k, e, \emb{v})S$, where $S$ is a substitution on $x_1 ,\ldots ,x_{n}$, and that $\sigma$ is an environment on $x_1,\ldots ,x_{n}$ such that $\interp$
    marries $\sigma$ and $S$. We have the following cases.
    \begin{description}
      \item[Case $e = v$:] Immediate.
      \item[Case $e = x$:] We have $x \in \dom(S) = \dom(\sigma)$. Because $c(k,e, \emb{v})
          = \mathsf{true}\wedge (\vemb{x} = \emb{v})$, we have $\interp(\vemb{x}S) = v$.
        Since $\interp$ marries $\sigma$ and $S$, we have that $\sigma(x)
          =\interp(\vemb{x}S)$. Because $\left< x, \sigma \right>\Downarrow_{\vset}
          \sigma(x)$ by \rname{Var}, we are done with this case.
      \item[Case $e = (\kwlet\ x = e_1\ \kwin\ e_2)$:] We have
        \[
          \interp \vDash \emap_s(k, e_1)S \wedge \emap_s(k + \n{e_1},e_2)S' \wedge (\emb{v} = \emap_\rho(k + \n{e_1}, e_2))S',
        \]
        where we set $S' = \{\emap_{\rho}(k, e_1)/ x\}S$. Thus both $\interp \vDash
          c(k, e_{1}, \emb{v_1})S$ and $\interp \vDash c(k + \n{e_1}, \emb{v_2})S'$,
        where we let $v_1 = \interp(\emap_\rho(k, e_1)S)$, and $v_2 =
          \interp(\emap_\rho(k + \n{e_1}, e_2)S')$. Thus, $\interp$ marries $\sigma'$
        with $S'$, where we set $\sigma' = \sigma[x \mapsto v_1]$. Therefore by two
        applications of the induction hypothesis, we have $\left< e_1,\sigma
          \right>\Downarrow_{\vset} v_1$ and $\left< e_2, \sigma'
          \right>\Downarrow_{\vset} v_2$. But then we can obtain by \rname[Let] that
        $\left< e, \sigma\right>\Downarrow_{\vset} v_2$. Because $\interp (\emb{v}) =
          \interp(\emb{v_2})$, we have that $v = v_2$ so we are done with this case.
      \item[Case $e = o(e_1 ,\ldots ,e_{n})$:] We have $\interp \vDash c(k + \n{e_1,\ldots
          ,e_{i-1}}, e_{i}, \emb{v_{i}})S$, for $i = 1,\ldots ,n$, where $v_{i} =
          \interp(\emap_\rho(k + \n{e_1 ,\ldots ,e_{i - 1}}, e_{i})S)$. By induction, we
        have $\left< e_{i}, \sigma \right>\Downarrow_{\vset} v_{i}$ for $i = 1,\ldots
          ,n$, which means that $\left< e, \sigma \right>\Downarrow_{\vset}
          \sem{o}(v_1,\ldots ,v_{n})$, using that $e$ is well-typed and \rname{Ops}.
        Because $\interp \vDash c(k, e, \emb{v})S$, we have that
        \[
          v = \interp(\emb{v}) = \interp(o (\emap_{\rho}(k, e_1) ,\ldots ,\emap_{\rho}(k + \n{e_1,\ldots ,e_{n-1}}, e_n))S) = \sem{o}(v_1,\ldots ,v_{n}),
        \]
        so we are done with this case.
      \item[Case $e = (\kwif\ e_1\ \kwthen\ e_2\ \kwelse\ e_3)$:] We have $\interp \vDash
          c(k, e_1,\emb{v_1} )S$, where $ v_1 = \interp(\emap_\rho(k, e_{1})S)$. Because
        $e$ is well-typed, we either have $v_1 = \mathsf{true}$ or $v_1 =
          \mathsf{false}$. Suppose the former. By induction, we have $\left<
          e_1,\sigma\right>\Downarrow_{\vset} \mathsf{true}$ and we have $\interp \vDash
          c(k + \n{e_1}, e_2, \emb{v})S$, using the definition of $\ite{}{}{}$ along with
        $\mu \vDash c(k, e, \emb{v})$. By induction, $\left< e_2,\sigma
          \right>\Downarrow_{\vset} v$. Thus $\left< e, \sigma \right>\Downarrow_{\vset}
          v$, using that $e$ is well-typed so that we can apply \rname{IfTrue}. The other
        case with $v_1 = \mathsf{false}$ is similar.
      \item[Case $e = \emark{a}{e_1}{e_2}$:] We have both $\interp \vDash c(k, e_1,
          \emb{v_1})S$ and $\interp \vDash c(k + \n{e_1}, e_2, \emb{v_2})S$, where we set
        $v_1 = \interp(\emap_{\rho}(k, e_1)S)$, and $v_2 =\interp(\emap_{\rho}(k +
            \n{e_1}, e_2)S)$. By induction, we have $\left< e_1,\sigma
          \right>\Downarrow_{\vset} v_{1}$ and $\left< e_2,\sigma \right>
          \Downarrow_{\vset} v_2$. Now let $r = \interp(y_{k + \n{e_1,e_2} + 1})$. Since
        $\interp \vDash c(k, e, v)S$, we also have $V_{a}([v_1, r]) = v_2$. But this
        means that $\left< e, \sigma \right> \Downarrow_{\vset} r$ by \rname{Mark}.
      \item[Case $e = \ediv{ e_1}{e_2}$:] We have $\interp \vDash c(k, e_{1},
          \emb{[r_1,r_1']})S$, and $\interp \vDash c(k + \n{e_1}, e_{2}, \emb{r_2})S$
        where we set $r_1 = \interp(\ell(\emap_{\rho}(k,e_1)S))$, $r_1' =
          \interp(r(\emap_{\rho}(k,e_1)S))$, and $r_2 = \interp(\emap_{\rho}(k +
            \n{e_1},e_2)S)$. By induction and $\interp$ being proper, we have that $\left<
          e_{1}, \sigma \right>\Downarrow_{\vset} [r_1,r_1']$ and $\left< e_2,\sigma
          \right>\Downarrow_{\vset} r_2$. Using that $\interp \vDash \emap_{s}(k, e)S$,
        we have that $r_1' \geq r_2\geq r_1$. Therefore, we can apply \rname{Div} to
        obtain $\left< e, \sigma \right>\Downarrow_{\vset}([r_1, r_2], [r_2, r_1'])$.
        Using that $\interp'$ is proper, we can argue similarly to the case of $e =
          o(e_1,\ldots ,e_{n})$ to obtain that $v = ([r_1, r_2], [r_2, r_1'])$, so we are
        done with this case.
    \end{description}
    The remaining cases are similar.
  \end{proof}

  \section{Implementation details}
  \label{app:impl}

  \paragraph{Constraint translation: if-then-else}
  Our implementation uses a slightly different translation for if-then-else,
  shown in \Cref{qsr-extra}. It is easy to see that these constraints are
  logically equivalent to the translation for if-then-else in \cref{qsr:a}, but
  we found that Z3 performs much better using the translation here.

  \begin{figure}[H]
    \begin{align*}
      \emap_Q(k, \kwif\ e_1\ \kwthen\ e_2\ \kwelse\ e_3)      & \d= k + \n{e_1, e_2 ,e_3} + 1                                                               \\
      \emap_{s}(k, \kwif\ e_1\ \kwthen\ e_2\ \kwelse\ e_3)    & \d= (((\emap_{\rho}(k, e_1) = \mathsf{true})\wedge \emap_{s}(k + \n{e_1}, e_2)              \\
                                                                                & \quad \quad\wedge (y_{k + \n{e_1, e_2,e_3} + 1} = \emap_{\rho}(k + \n{e_1}, e_2))           \\
                                                                                & \ \quad \vee ((\emap_{\rho}(k, e_1) = \mathsf{false})\wedge \emap_{s}(k + \n{e_1,e_2}, e_3) \\
                                                                                & \quad \quad \wedge (y_{k + \n{e_1, e_2,e_3} + 1}= \emap_{\rho}(k + \n{e_1, e_2}, e_3)))     \\
                                                                                & \ \quad\wedge \emap_{s}(k, e_1)                                                             \\
      \emap_{\rho} (k,\kwif\ e_1\ \kwthen\ e_2\ \kwelse\ e_3) & \d=  y_{k + \n{e_1 , e_2, e_3}}                                                             \\ \\
    \end{align*}
    \caption{Constraint translation.}
    \label{qsr-extra}
  \end{figure}

  \paragraph{Axioms on valuations.} 
  We used the following axioms for valuations in our implementation:
  \begin{enumerate}
    \item $V(\left[ 0,1 \right]) = 1$
    \item $V(I) \geq 0$
    \item $(\ell(I_2) = r(I_1)) \Rightarrow V(I_1)+ V(I_2) = V([\ell (I_1), r(I_2)])$
    \item $I_2 \subseteq I_1 \Rightarrow V(I_2) \leq V(I_1)$
    \item $V(I) \leq 1$
    \item $(I =  \{r\}) \Rightarrow V(I)= 0$
    \item $\ell(I_1)\geq \ell(I_2), V(I_1) \geq V(I_2) \Rightarrow r(I_1)\geq r(I_2)$
    \item $r(I_1)\geq r(I_2), V(I_2) \geq V(I_1) \Rightarrow \ell(I_1)\geq \ell(I_2)$
  \end{enumerate}
  for any intervals $I,I_1,I_2\subseteq \left[ 0,1 \right]$. The first three
  axioms correspond to the first three assumptions in Section~\ref{sec:motiv}.
  The continuity assumption is already built into the constraint generated for
  $\kwmark$; indeed, all protocols in the Robertson-Webb model require this
  assumption on valuations. The remaining assumptions are general facts about
  valuations; while they are derivable from the first three axioms, we found that
  including these assumptions was useful for Z3.

  \section{A Custom Constraint Translation for Verifying Progress}

  We've seen in \Cref{sec:constraints} how to verify properties that concern the
  output of protocols. However, what about properties of protocols that don't
  concern the output? One such property is progress, that is, a protocol will
  always step to a value, assuming it is well-typed. It could be the case that a
  protocol is envy-free in the sense that whenever it steps to an allocation,
  that allocation is envy-free, yet the protocol may get stuck. Certainly this
  kind of behavior is not ideal, so that progress is valuable to verify.

  Completeness from \Cref{sec:constraints} could be employed to show this, however that would require showing that $c(e, \ret)$ is satisfiable in $\M_{\vset}$ for every set of valuation functions $\vset$. This approach does not as easily lend itself to SMT solvers. In this section, we propose another translation that lends itself more easily to SMT solvers.

  To be more clear, progress is the property that a protocol will always step to
  a value. This means that whenever $e = \emark{a}{e_1}{e_2}$ and $\left< e_1,
    \sigma \right> \Downarrow v_1$, $\left< e_2, \sigma \right>\Downarrow v_2$, we
  require that there exists $r\in [0,1]$ such that $V_{a}([v_1, r]) = v_2$, and
  whenever $e = \ediv{e_1}{e_2}$, and $\left< e_1, \sigma \right>\Downarrow [r_1,
      r_1']$, $\left< e_2, \sigma \right>\Downarrow r_2$, it must be the case that
  $r_1 \leq r_2 \leq r_1'$. In light of \Cref{prop:soundness}, if we can show
  that $c(k, (e_1, e_2), (v_1,v_2))$ implies that $\v_{a}([v_1, 1])\geq v_2$, and
  that $c(k, (e_1,e_2), (r_1,r_2))$ implies that $r_1' \geq r_2 \geq r_1$, then
  we are able to show that a protocol always has progress. However, branch
  conditions may also help show these facts. These considerations motivate the
  following definition.

  \begin{figure}
    \begin{align*}
      m_{a}(k, e_1, e_2, B, S) & \d=\forall \ret. \exists y_1 ,\ldots ,y_{k + \n{e_1 ,e_2}}.(c(k, (e_1,e_2), \ret)S\wedge B) \Rightarrow (V_{a} (\left[ \pi_{1}\ret, 1 \right]) \geq \pi_{2}\ret)))  \\
      d(k, e_1, e_2, B, S)     & \d=\forall \ret. \exists y_1 ,\ldots ,y_{k + \n{e_1 ,e_2}}.( c(k ,(e_1,e_2), \ret)S\wedge B) \Rightarrow (\ell( \pi_{1}\ret) \leq \pi_{2}\ret \leq r(\pi_{1}\ret)))
    \end{align*}
    \begin{align*}
      \ef (k, \emark{a}{e_1}{e_2}, B, S)                               & \d= \ef(k, e_1, B, S)\wedge \ef(k, e_2, B, S) \wedge m_{a}(k,e_1, e_2, B, S)                                       \\
      \ef (k, \ediv{e_1}{e_2}, B, S)                                 & \d= \ef(k, e_1, B, S)\wedge \ef(k, e_2, B, S) \wedge d(k, e_1, e_2,  B, S)                                         \\
      \ef(k, \kwif\ e_1\ \kwthen\ e_2\  \kwelse\ e_3,  B, S) & \d= \ef (k, e_1, B, S)                                                                                             \\
                                                                               & \quad\wedge \ef (k + \n{e_1}, e_2, (\emap_{\rho}(k, e_1) = \kwtrue)S\wedge \emap_{s}(k, e_1)S\wedge B, S)    \\
                                                                               & \quad\wedge \ef (k + \n{e_1}, e_3, (\emap_{\rho}(k, e_1) = \kwfalse)S\wedge \emap_s(k, e_1)S \wedge B, S)    \\
      \ef(k, \kwlet\ x = e_1\ \kwin\ e_2, B, S)                    & \d= \ef(k, e_1, B, S) \wedge \ef(k + \n{e_1}, e_2, B\wedge \emap_{s}(k, e_1)S, \{\emap_{\rho}(k, e_1)/\vemb{x}\}S) \\
      \ef(k, v, B, S)                                                          & \d= \kwtrue                                                                                                  \\
      \ef(k, x, B, S)                                                          & \d= \kwtrue                                                                                                  \\
      \ef(e(e_1 ,\ldots ,e_{n}),  B, S)                                        & \d= \ef(k, e_1, B, S)\wedge \cdots \wedge \ef(k, e_n, B, S) \quad \text{for all remaining expressions.}
    \end{align*}
    \caption{Constraint translation: progress.}
    \label{def:errfree}
  \end{figure}

  Let $\ef :\mathbb{N}\times \mathtt{Exp} \times \f \times \mathbf{S}\to \f $ be as given on
  expressions in Figure~\ref{def:errfree}, where $\mathbf{S}$ is the set of substitutions. We give a high level explanation. We
  would like it to be the case that if $\ef(k, e, B, S)$ is interpreted to be
  true, then $e$ always steps to a value. This is why for all expressions besides
  mark, divide, and if then else, $\ef$ is just passed through its
  subexpressions. $B$ is meant to be thought of as a formula keeping track of
  branch conditions and previous side conditions in the protocol, which is why
  $B$ only gets updated when passing to subexpressions for conditionals and the
  let binding. The substitution $S$ is required for properly keeping track of
  bound program variables, in a similar way that they are used in
  \Cref{sec:constraints}. We would like to keep just enough facts in order to
  always prove progress. The following example may be illuminating.

  \paragraph{Example}
  Consider again Cut-choose, from Figure~\ref{proto:cc}. We have
  \begin{align*}
     & \ef(0, e, \mathsf{true}) = \forall \ret. (\ret = (0,1/2)) \Rightarrow (\v_{1}( [\pi_1 \ret, 1])\geq \pi_2\ret)\wedge                                                     \\
     & \forall \ret. (\exists y_1. (\v_{1}([0,y_1]) = 1/2)\wedge (\v_{1}([0,1])\geq 1/2)\wedge (\ret = y_1)) \Rightarrow r(\pi_1(\ret))\geq \pi_2(\ret) \geq \ell(\pi_1(\ret)).
  \end{align*}
  where we just omit all conjunctions with $\mathsf{true}$. The first line corresponds to the mark subexpression, while the second line corresponds to the divide subexpression. It is clear from inspection that $\ef(0 , e, \mathsf{true} )$ is satisfiable by any proper interpretation.
  \\ \\

  We now ramp up to showing that $\ef$ achieves its purpose through some helpful
  technical facts.
  \begin{lemma}
    For any formula $B$ and substitution $S$ such that $\fv(B)\cup \fv(S)\subseteq \{y_1 ,\ldots ,y_{k}\}$,
    \[
        \fv(\ef(k', e , B, S))\cap \mathcal{Y} = \emptyset
    \]
    if $k' \geq k$. \label{lem:yina}
  \end{lemma}
  \begin{proof}
    Induction on the structure of $e$.
  \end{proof}
  \begin{lemma}
    Let $e$ be any expression, $B$ any formula, and $S$ any substitution such that $\fv(e)\subseteq \dom(S)$ and $\fv(B)\mathcal{X} = \emptyset$. Then
    $\fv(\ef(k, e, B, S))\cap \mathcal{X} = \emptyset$.
    \label{lem:xina}
  \end{lemma}
  \begin{proof}
    Induction on the structure of $e$.
  \end{proof}
  In light of these facts, we now state and prove our desired result.
  \begin{proposition}
    \label{prop:errf}
    Let $\left< e,\sigma \right>$ be a configuration such that $\fv(e)\subseteq \dom(\sigma)$.
    If there is a formula $B$, substitution $S$, and interpretation $\interp \in \M_{\vset}$ and natural number $k$ such that
    \begin{enumerate}
      \item $\fv(B)\subseteq \{y_1,\ldots ,y_{k}\}$
      \item $k \geq m(S)$
      \item $\interp$ marries $\sigma$ and $S$
      \item $\interp \vDash \ef (k, e, B, S) \wedge B$
    \end{enumerate}
    Then $\left< e,\sigma \right>\Downarrow_{\vset} v$ for some $v\in \val$.
  \end{proposition}
  \begin{proof}
    We proceed by induction on the sub-expression relation. \\
    In the case that either $e = v$ or $e = x$, we are immediately done. \\ \\
    Suppose that $e = \emark{a}{e_1}{e_2}$. By definition of $\ef$, we
    have that both $\interp \vDash \ef(k , e_1, B, S)$ and $\interp \vDash \ef(k
      ,e_2, B, S)$. By induction, there exists $v_1,v_2$ such that $\left< e_1,\sigma
      \right>\Downarrow_{\vset} v_1$, and $\left< e_2,\sigma \right> \Downarrow_{\vset} v_2$. We then
    have $\left< (e_1,e_2), \sigma\right>\Downarrow_{\vset} (v_1,v_2)$. By
    \Cref{prop:soundness}, we have that
    \[
      \interp' \vDash c(k , (e_1,e_2), (v_1, v_2))S
    \]
    for some proper interpretation $\interp'$ that agrees with $\interp$ up to $k$.
    Thus we can say that
    \[
      \interp \vDash \exists y_1 ,\ldots ,y_{k + \n{e_1,e_2}}. c(k , (e_1,e_2), (v_1, v_2))S)
    \]

    But since $\interp \vDash B$, we also have that $\interp' \vDash B$, using
    assumption (1) and that $\interp'$ agrees with $\interp$ up to $k$, hence
    \[
      \interp' \vDash (c(k , (e_1,e_2), (v_1, v_2))S \wedge B.
    \]
    By \Cref{lem:yinc} and assumption (2) and the above, we have
    \[
      \fv(c(k , (e_1,e_2), (v_1, v_2))S\wedge B) \subseteq \{y_1 ,\ldots ,y_{k + \n{e_1, e_2}}\},
    \]
    so we also obtain
    \[
      \interp \vDash \exists y_1 ,\ldots ,y_{k + \n{e_1,e_2}}.
      c(k , (e_1,e_2), (v_1, v_2))S\wedge B.
    \]
    Thus because $\interp \vDash m_{a}(k, e_1,e_2, B,S)$, taking $\ret$ to be
    $\emb{(v_1,v_2)}$, we obtain $\interp \vDash \v_{a} (\left[ \emb{v_1}, 1
        \right]) \geq \emb{v_2}$. This means that there is some $r \in \mathbb{R}$ such
    that $V_{a}(\left[ v_1,r \right]) = v_2$. We can then conclude that $\left<
      e,\sigma \right>\Downarrow_{\vset} r$.\\ \\
      Now if $e = \ediv{e_1}{e_2}$, quite a similar argument can be made.
    \\ \\
    Suppose that $e = \kwif\ e_1\ \kwthen\ e_2\ \kwelse\ e_3$.
    Then we have that
    \[
      \interp \vDash \ef(k, e_1, B, S),
    \]
    \[
      \interp \vDash \ef (k + \n{e_1}, e_2, (\emap_{\rho}(k , e_1) = \kwtrue)S\wedge \emap_{s}(k , e_1)S\wedge B, S)
    \]
    and
    \[
      \interp \vDash \ef (k + \n{e_1}, e_3, (\emap_{\rho}(k , e_1) = \kwfalse)S\wedge \emap_s(k , e_1)S \wedge B, S).
    \]
    Immediately by induction, $\left< e_1, \sigma\right> \Downarrow_{\vset} \mathsf{true}$
    or $\left< e_1, \sigma \right> \Downarrow_{\vset} \mathsf{false}$. Suppose the former
    is true. Then by \Cref{prop:soundness}, we have that there exists $\interp'$
    such that $\interp' \vDash c(k , e_1, \emb{\mathsf{true}})S$ and $\interp'$
    agrees with $\interp$ up to $k$. This means that $\interp' \vDash B'$, where
    \[
      B' = (\emap_{\rho}(k , e_1) = \kwtrue)S\wedge \emap_{s}(k , e_1)S\wedge B.
    \]
    Notice that $\fv(B')\cap \mathcal{X} \subseteq \dom(\sigma)$ by \Cref{lem:xinc}
    and that $\fv(B')\cap \mathcal{Y}\subseteq \{y_{1} ,\ldots ,y_{k + \n{e_1}}\}$
    by \Cref{lem:yinc} and by assumption on $B$. Because
    \[
      \fv(\ef (k + \n{e_1}, e_2, (\emap_{\rho}(k , e_1) = \kwtrue)S\wedge \emap_{s}(k , e_1)S\wedge B,S)) = \emptyset,
    \]
    by \Cref{lem:xina} and \Cref{lem:yina}, we also have that
    \[
      \interp' \vDash \ef (k + \n{e_1}, e_2, (\emap_{\rho}(k , e_1) = \kwtrue)S\wedge \emap_{s}(k , e_1)S\wedge B, S),
    \]
    as both $\interp$ and $\interp'$ are proper. Thus we can apply induction to
    obtain that $\left< e_2, \sigma \right> \Downarrow_{\vset} v_2$ for some $v_2$. This
    means that $\left< e, \sigma \right>\Downarrow_{\vset} v_2$. The latter case is argued
    similarly.\\ \\
    Suppose $e = (\kwlet\ x= e_1\ \kwin\ e_2)$. Then $\interp \vDash
      \ef(k , e_1, B, S)$ and $\interp \vDash \ef(k + \n{e_1}, e_2, B', S')$, where
    we set $B' = B\wedge \emap_{s}(k,e_1)S$ and $S'= \{\emap_{\rho}(k,e_1)/ x\} S$.
    We can apply induction immediately to obtain that $\left< e_1,\sigma
      \right>\Downarrow_{\vset} v_1$ for some $v_1$. By \Cref{prop:soundness}, there exists a
    proper interpretation $\interp'$ such that
    \[
      \interp' \vDash c(k, e_1, \emb{v_1})S,
    \]
    and $\interp'$ agrees with $\interp$ up to $k$. Expanding the constraint of
    $e_1$, we see that $\interp'(\emap_\rho(k, e_1)S) = v_1$ so that $\interp'$
    marries $\sigma[x \mapsto v_1]$ and $S'$. Now by \Cref{lem:yina} and
    \Cref{lem:xina},
    \[
      \fv(\ef(k  + \n{e_1}, e_2, B', S')) = \emptyset,
    \]
    so we also have
    \[
      \interp' \vDash \ef(k  + \n{e_1}, e_2, B', S').
    \]
    Recalling that $\interp'$ agrees with $\interp$ up to $k$ and $\fv(B)\cap
      \mathcal{Y} \subseteq \{y_1,\ldots ,y_{k}\}$, we obtain that $\interp' \vDash
      B'$ as well. Therefore we can apply induction again to obtain $\left< e_2,
      \sigma[x \mapsto v_1] \right> \Downarrow_{\vset} v_2$ for some $v_2 \in \val$. This
    means that $\left< e, \sigma \right>\Downarrow_{\vset} v_2$.
  \end{proof}

  The above proposition has many conditions. As we have seen in earlier, it
  becomes much simpler when an expression has no free variables.
  \begin{corollary}
    Let $e$ be an expression with no free variables. If there exists $\interp \in \M_{\vset}$ such that
    \[
      \interp \vDash \ef(0, e, \mathsf{true},\varepsilon),
    \]
    then $\left< e,\varepsilon \right>\Downarrow_{\vset} v$ for some $v \in \val$.
    \label{cor:af}
  \end{corollary}
  We can apply this result as follows. Given an expression with no free variables, $\ef(0,e, \mathsf{true}, \varepsilon)$, also has no free variables by \Cref{lem:xina} and \Cref{lem:yina}. Thus $\ef(0,e,\mathsf{true},\varepsilon)$ is satisfiable in $\M_{\vset}$ if and only if it is valid in $\M_{\vset}$. So if we can show $\vDash \ef(0,e,\mathsf{true},\varepsilon)$, then we have shown that $\left< e,\varepsilon \right>\Downarrow_{\vset} v$ for some $v \in \val$ for any $\vset$. SMT solvers can be employed to check $\vDash \ef(0,e,\mathsf{true},\varepsilon)$ in a similar way described in \Cref{sec:impl}.

\section{Selfridge-Conway-Surplus}
\label{app:sc-plus}

The Selfridge-Conway-Surplus protocol follows the Selfridge-Conway protocol but
stops early, disposing the $\mathit{Rest}$ interval instead of dividing it among
the agents.  \Cref{code:sc_surplus} presents the code for this protocol, and
\Cref{fig:sc_surplus} shows one possible allocation.

\begin{figure}[h]
  \centering
  \captionsetup{justification=centering}
  \begin{minipage}{0.49\linewidth}
    \[
      \small\begin{array}{l}
        \kw{SelfridgeConwaySurplus} :                                           \\
        \quad\kwlet\ (I_1, I_1') = \ediv{\kwcake}{\emark{1}{0}{1 / 3}}\ \kwin\  \\
        \quad\kwlet\ (I_2, I_3) = \ediv{I_1'}{\emark{1}{0}{2 / 3}}\ \kwin\      \\
        \quad\kwlet\ (A, B, C) = \kwsort_2(I_1, I_2, I_3)\ \kwin\               \\
        \quad\kwif\ \eeval{2}{A}\ = \eeval{2}{B}\ \kwthen\                      \\
        \quad\quad \kwlet\ (A, B, C) = \kwsort_3(A, B, C)\ \kwin\               \\
        \quad\quad \kwlet\ (B, C) = \kwsort_2(B, C)\ \kwin\                     \\
        \quad\quad \kwalloc(C, B, A)                                            \\
        \quad\kwelse\                                                           \\
        \quad\quad \kwlet\ m = \emark{2}{\kwlt\ A}{\eeval{2}{B}}\	\kwin\        \\
        \quad\quad \kwlet\ (\mathit{Trim}, \mathit{Rest}) = \ediv{A}{m}\	\kwin\ \\
        \quad\quad\kwlet\ (A, B, C) = \kwsort_3(\mathit{Trim}, B, C)\ \kwin\    \\
        \quad\quad\kwif\ A \neq \mathit{Trim}\ \kwthen\                         \\
        \quad\quad\quad\kwif\ B = \mathit{Trim}\ \kwthen\                       \\
        \quad\quad\quad\quad \kwalloc(C, B, A,)                                 \\
        \quad\quad\quad \kwelse\                                                \\
        \quad\quad\quad\quad \kwalloc(B, C, A)                                  \\
        \quad\quad \kwelse\                                                     \\
        \quad\quad\quad \kwlet\ (B, C) = \kwsort_2(B, C)\ \kwin\                \\
        \quad\quad\quad \kwalloc(C, B, A)                                       \\
      \end{array}\normalsize
    \]
    \caption{SelfridgeConwaySurplus in \SYSTEM.}
    \label{code:sc_surplus}

  \end{minipage}\hfill
  \begin{minipage}{0.49\linewidth}
    \includegraphics[width=\linewidth]{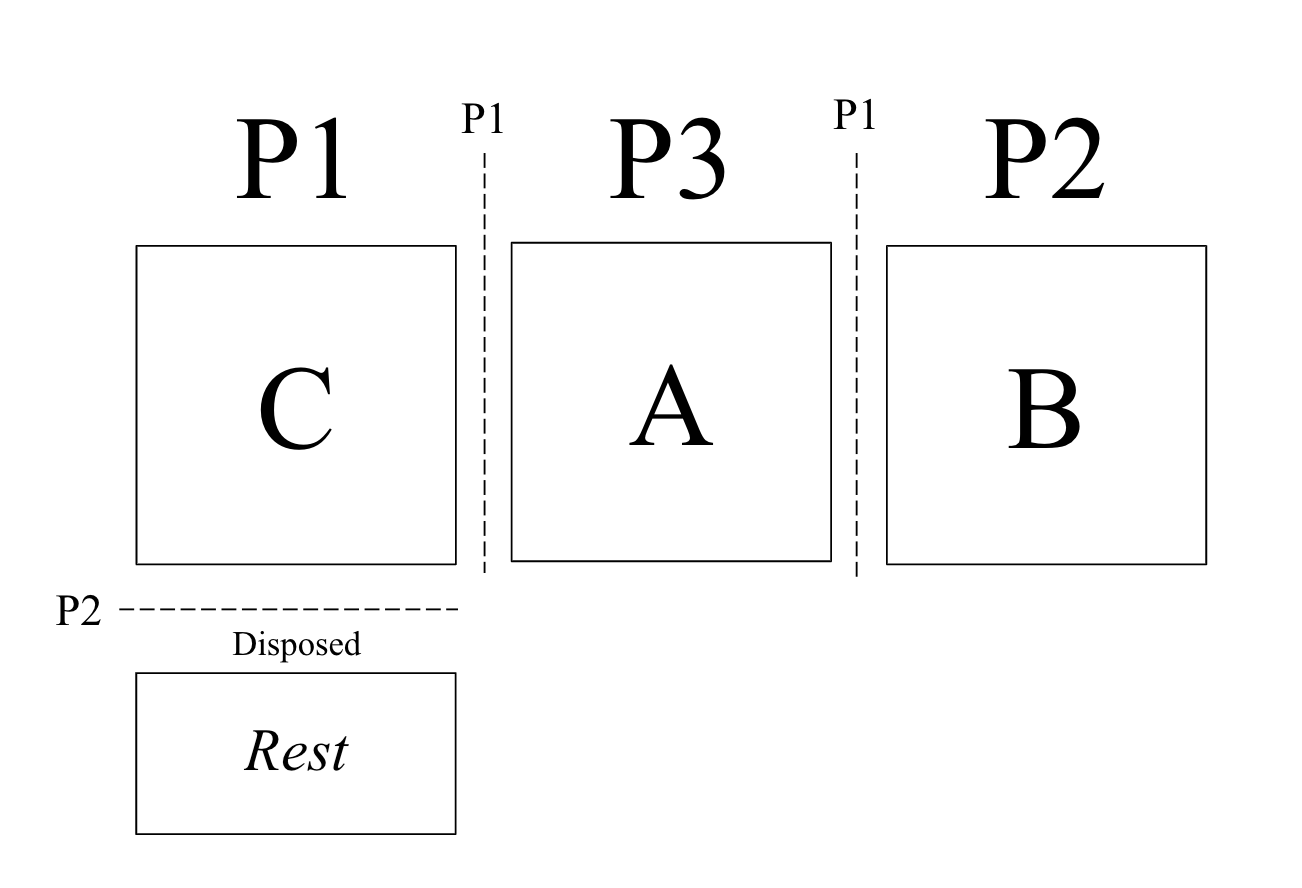}
    \caption{Possible allocation from SelfridgeConwaySurplus.}
    \label{fig:sc_surplus}
  \end{minipage}
\end{figure}
\fi
\end{document}